\documentclass[12pt]{article}
\usepackage{amsmath}
\usepackage{amssymb}
\usepackage[unicode=true,pdfusetitle,
 bookmarks=true,bookmarksnumbered=false,bookmarksopen=false,
 breaklinks=true,pdfborder={0 0 1},
colorlinks=false]
 {hyperref}
\usepackage{breakurl}
\usepackage{slashed}
\usepackage{axodraw,3pt}
\usepackage{epsfig}
\usepackage{bbm}
\usepackage{bm}
\usepackage{color}

\usepackage{mathrsfs}
\usepackage{dsfont}
\renewcommand{\marginpar}[1]{}

\usepackage{appendix}

\usepackage{exscale}

\def\a0size{6}

\newcommand{\lsi}{\raise0.3ex\hbox{$<$\kern-0.75em\raise-1.1ex\hbox{$\sim$}}}
\newcommand{\gsi}{\raise0.3ex\hbox{$>$\kern-0.75em\raise-1.1ex\hbox{$\sim$}}}
\newcommand{\lsim}{\mathop{\lsi}}
\newcommand{\gsim}{\mathop{\gsi}}
\newcommand{\sumint}[1]{{\hbox{$\sum$}\!\!\!\!\!\!\!\int\,}
_{\!\!\!\!\!\!\!\!\!\!\raise-1.2ex\hbox{$\scriptstyle{#1}$}}}

\renewcommand{\vec}[1]{{\bf #1}}
\newcommand{\cmatrix}{{\cal C}}

\newcommand{\cc} {\textsf{C}}
\newcommand{\pa} {\textsf{P}}
\newcommand{\tr} {\textsf{T}}

\newcommand{\ord}{O}
\newcommand{\lepton}{\ell}

\newcommand{\pv}{{\rm P.V.}} 
\newcommand{\teuc}{{\rm T}} 
\newcommand{\Nweak}{N_{\rm w}}

\newcommand*\xbar[1]{%
  \hbox{%
    \vbox{%
      \hrule height 0.5pt 
      \kern0.5ex
      \hbox{%
        \kern-0.1em
        \ensuremath{#1}%
        \kern-0.1em
      }%
    }%
  }%
} 

\allowdisplaybreaks[1]

\makeatletter
\newsavebox\myboxA
\newsavebox\myboxB
\newlength\mylenA

\newcommand*\xoverline[2][0.75]{%
    \sbox{\myboxA}{$\m@th#2$}%
    \setbox\myboxB\null
    \ht\myboxB=\ht\myboxA%
    \dp\myboxB=\dp\myboxA%
    \wd\myboxB=#1\wd\myboxA
    \sbox\myboxB{$\m@th\overline{\copy\myboxB}$}
    \setlength\mylenA{\the\wd\myboxA}
    \addtolength\mylenA{-\the\wd\myboxB}%
    \ifdim\wd\myboxB<\wd\myboxA%
       \rlap{\hskip 0.5\mylenA\usebox\myboxB}{\usebox\myboxA}%
    \else
        \hskip -0.5\mylenA\rlap{\usebox\myboxA}{\hskip 0.5\mylenA\usebox\myboxB}%
    \fi}
\makeatother

\begin{document}
\setlength{\baselineskip}{0.6cm} \global\long\def\figysize{16.0cm}
 \global\long\def\figtopspace{\vspace*{-1.5cm}}
 \global\long\def\figbottomspace{\vspace*{-5.0cm}}

\global\long\def\theequation{\thesection.\arabic{equation}}
 \newcounter{saveeqn}
\newcommand{\alphaeqn}{\refstepcounter{equation}\setcounter{saveeqn}{\value{equation}}%
\setcounter{equation}{0}%
\renewcommand{\theequation}{%
        \mbox{\thesection.\arabic{saveeqn}\alph{equation}}}}%
\newcommand{\reseteqn}{\setcounter{equation}{\value{saveeqn}}%
\renewcommand{\theequation}{\thesection.\arabic{equation}}}

\makeatletter \@addtoreset{equation}{section} \makeatother
\renewcommand{\theequation}{\arabic{section}.\arabic{equation}}



\begin{centering}

  \vfill

  \textbf{\Large{{Lepton asymmetry rate from quantum field theory: 
      \\[2.5mm] 
      NLO in the hierarchical limit
  }}}

\vspace*{.6cm}

D.~~B\"odeker
\footnote{bodeker@physik.uni-bielefeld.de%
} and M.~Sangel
\footnote{msangel@physik.uni-bielefeld.de%
} 

\vspace*{.6cm}

{\em Fakult\"at f\"ur Physik, Universit\"at Bielefeld, 33501 Bielefeld,
Germany } 

\vspace{10mm}

\textbf{Abstract} 

\end{centering}

\vspace{5mm}
\noindent
The rates for generating a matter-antimatter asymmetry in extensions of
the Standard Model (SM) containing right-handed neutrinos are the most
interesting and least trivial co\-efficients in the rate equations for
baryogenesis through thermal leptogenesis.  We obtain a relation of
these rates to finite-temperature real-time correlation functions,
similar to the Kubo formulas for transport coefficients.  Then we
consider the case of hie\-rarchical masses for the sterile neutrinos.
At leading order in their Yukawa couplings we find a simple master
formula which relates the rates to a single finite temperature
three-point spectral function. It is valid to all orders in $ g $,
where $ g $ denotes a SM gauge or quark Yukawa coupling.  
We use it to compute the rate for 
generating a matter-antimatter asymmetry at next-to-leading order in 
$ g $ 
in the non-relativistic regime. The corrections
are of order $ g ^ 2 $, and they amount to  4\% or less.

\noindent \vspace{0.5cm}

\noindent 
\vspace{0.3cm}

\noindent \vfill
 \vfill

\newpage

\section{Introduction \label{sec:intro}}

The Standard Model of particle physics has been very successfully
tested at high energies.  It can, however, neither explain neutrino
oscillations, nor the matter-antimatter asymmetry of the
Universe. These two open problems may have an elegant common
solution. By extending the Standard Model by right-handed, or sterile neutrinos
one can give masses to neutrinos which can cause them to
oscillate. Their Yukawa couplings introduce a new source of $ CP $
violation. These couplings are very small, so that 
sterile
neutrinos easily deviate from thermal equilibrium in the early
Universe. Electroweak sphalerons
rapidly violate baryon plus lepton number at temperatures $ T \gsim 130 $GeV.
Then all three Sakharov conditions are satisfied and the
baryon asymmetry of the Universe can be
generated~\cite{Fukugita:1986hr}.  This is referred to as
baryogenesis through leptogenesis.

Many aspects and scenarios of leptogenesis have been studied, covering  a
vast range of sterile-neutrino masses (for reviews see
e.g.~\cite{davidson-review,buchmuller-review}).
Originally leptogenesis was formulated using the Boltzmann equation,
augmented with additional prescriptions, to overcome inconsistencies
or to account for medium effects. The Boltzmann equation already
contains a set of implicit assumptions and approximations. In order to
eliminate ambiguities, and to assess the accuracy of the calculation
of the baryon asymmetry one has to start from first principles.  This
has led various authors to start from Kadanoff-Baym or similar
equations for Green's functions (cf.\
Refs.~\cite{beneke-density,anisimov-quantum,garny-resonant} for recent
work and references).  This way ambiguities have been clarified for
resonant leptogenesis and a computation of flavor effects was
possible~\cite{beneke-flavoured}.  In order to assess the theoretical
error of leptogenesis calculations one has to identify appropriate
expansion parameters and then compute corrections in this expansion.
Being exact, the Kadanoff-Baym equations are difficult to handle, and
it is particularly hard to compute corrections.

Here we follow another first-principle
approach~\cite{Bodeker:2013qaa,bodeker-washout,occupation}, in which
radiative corrections have already been successfully included.  Right from the
start it makes use of two key features of leptogenesis: ($i$) Most
degrees of freedom in the hot plasma are kept in thermal equilibrium
by Standard Model processes with interaction rates much larger than
the Hubble rate, the so called {\it spectator processes}.  ($ii$)
There is a separation of time scales; the quantities which deviate
from equilibrium evolve on scales of order of the Hubble time or even
larger times, i.e., much more slowly than the spectator
processes.
Under these conditions
the non-equilibrium state is
specified by the temperature, and by the values of the slow
variables.  In particular, these quantities determine the time
evolution of the slowly changing 
variables. For sufficiently small deviations $ y
_ a $ from equilibrium the equations can be linearized. Then,
in the absence of expansion, the non-equilibrium process
can be  described by the effective classical\footnote{Since
the time scale on which the $ y _ \alpha  $ evolve is larger  than
the inverse temperature, the $ y _ \alpha  $ behave classically (see, e.g., 
Ref.~\cite{Landau}, \S 110).} 
kinetic equations  (cf.\ Ref.~\cite{Landau})
\begin{equation}
  \dot y_{a  }=-\gamma_{a  b  }\, y_{b  }
  \label{eom} 
  .
\end{equation}
The real coefficients $ \gamma  _ { ab } $ only depend on the temperature and
encode the effect of the spectator processes.\footnote{They
  can also depend on the values of conserved or practically conserved
  charges but these are usually assumed to vanish.}  They have to 
be determined from the underlying microscopic theory. 
One arrives at relations which are quite similar
to the Kubo formulas for transport coefficients.  In these relations
the $ \gamma _ { ab } $ are written in terms of equal time correlation
functions, so-called susceptibilities, and of unequal-time correlation
functions of the slowly changing variables $ y _  \alpha  $. These
correlation functions are evaluated in an {\it equilibrium}
system. This way one can relate the $ \gamma _ { ab } $ to objects which
can be computed in finite temperature field theory.
Then the computation of the baryon asymmetry
proceeds in two separate steps. First one computes the coefficients in the rate
equations in (\ref{kineq1}), (\ref{kineq2}) using quantum field theory
for equilibrium systems. The non-equilibrium problem is then treated by
solving the rate equations. 
This is  an enormous simplification 
compared to other approaches  where
the computation of radiative corrections appears to be prohibitively 
difficult. 

What the slow variables are depends on the model parameters and 
on the relevant temperature range. In the
limit of hierarchical sterile neutrino-masses $M_1\ll M_{I\not=1}$,
only the lightest sterile neutrinos $ N _ 1 $ are present in the
plasma for $ T \sim M _ 1 $. 
Then the slow variables are the (spatially homogeneous) $ N _
1 $-phase-space density $f_{\vec{k}}$, as well as global charges $ X _
a $ such as $L-B$ where $ L $ and $ B $ are lepton and baryon number.
Both types of quantities are conserved by Standard Model interactions,
their conservation is violated only by the sterile-neutrino Yukawa
interactions. The corresponding rates are small due to the smallness
of the Yukawa couplings.  In thermal leptogenesis these rates become
similar in size to the Hubble rate.  If we assume that during
leptogenesis the deviation of the sterile neutrino phase-space density
from equilibrium
\begin{equation}
  \delta f_{\vec{k}}
  \equiv 
  f_{\vec{k}}-f_{\vec{k}}^{\rm eq} 
  ,
\end{equation}
and the values of the charges $X_a$
are sufficiently 
small, the system is described by linear kinetic equations of the 
form\footnote{We consider a finite volume $V$
so that the momenta $\vec{k}$ are discrete. Summation over
indices appearing twice is understood.}
\begin{align}
  D_{t} f_{\vec{k}} 
  & =
  - \gamma_{ \vec{k}  \vec q } \, \delta f_{\vec{q}}
   - \gamma_{\vec{k}a} \, X_a
   ,
  \label{kineq1}\\
  D_{t}X_{a} 
  & =
  - \gamma_{a\vec{k}} \, \delta  f_{\vec{k}}
   - \gamma_{ab} \, X_{b}
  ,
  \label{kineq2}
\end{align}
where $D_{t}$ is the time derivative which takes into account the
expansion of the Universe.  

Kubo-type relations for the washout rates $\gamma_{ab}$ in
(\ref{kineq2}), and the sterile-neutrino equilibration rate $\gamma_{
  \vec k \vec q}$ in (\ref{kineq1}) were obtained in
Refs.~\cite{bodeker-washout} and \cite{occupation}, respectively.
These are valid to leading order in the sterile-neutrino Yukawa
couplings, and to {\it all} orders in Standard Model couplings.\footnote{To
systematically include higher orders in the sterile-neutrino Yukawa
interactions, one would also have to consider higher derivative terms
in Eqs.~(\ref{kineq1}), (\ref{kineq2}) which are similarly
parametrically suppressed.}  The next-to-leading order (NLO) Standard
Model corrections to $ \gamma _ { \vec k \vec q } $ are known in the
regime $ T \ll M _ 1 $ \cite{Laine:2011pq}, where the $ N _ 1 $ are
non-relativistic, and in the relativistic regime $ T \sim M _ 1 $
\cite{Laine:2013lka}.  In both cases the NLO is of order $ g ^ 2 $,
where $ g $ denotes some generic Standard Model coupling. In the
ultrarelativistic regime $ T \gsim M _ 1 /g $ even the leading order
(LO) result is quite involved, it was calculated in
Refs.~\cite{anisimov,besak}.  The leading corrections to the washout
rates $ \gamma _ { ab } $ are only suppressed by a single power of $ g
$; the order $ g $ and order $ g ^ 2 $ corrections were computed in
Refs.~\cite{bodeker-washout,Bodeker:2015zda}.

In this paper we consider the $CP$ violating rates $\gamma_{a\vec{k}}$
and $\gamma_{\vec{k}a}$ in (\ref{kineq1}) and (\ref{kineq2}).  We
obtain the master formulas (\ref{mastereff}) and (\ref{master2}) which
relate them to a single three-point spectral function of Standard
Model fields. These are valid to LO in the sterile-neutrino Yukawa
interaction, and to all orders in the Standard Model gauge couplings
and the quark Yukawa couplings. The small $CP$ violation of the
Standard Model is neglected, together with the charged lepton
Yukawa-interactions.  We evaluate our master formulas in the regimes $
T \ll M_1$ and $ T \sim M _ 1 $ at leading order in Standard Model
couplings.  Then we perform the first step of the order $g^2$
calculation of the $CP$ violating lepton asymmetry rate
$\gamma_{a\vec{k}}$ in the regime $ T \ll M_1$ by computing the zero
temperature contribution, which is the leading term in the
low-temperature expansion.

This paper is organized as follows. In Sec.~\ref{rates} we slightly
generalize the method of Ref.~\cite{bodeker-washout} which allows us
to obtain Kubo-type relations for $ CP $ violating rates.  In
Sec.~\ref{general} we derive the master formulas for these rates, and
in Sec.~\ref{lepton} we demonstrate how they reproduce the
leading-order lepton asymmetry rate for $ T \ll M _ 1 $ and
$ T \sim M _ 1 $. Then in Sec.~\ref{zero} we
compute the order $g^2$ corrections to the asymmetry rate at zero
temperature. We summarize in Sec.~\ref{s:summary}. Appendix~\ref{a:3pt}
contains a derivation of a spectral representation for arbitrary
thermal three-point functions of bosonic or fermionic operators.
Implications of discrete symmetries for spectral functions are the
subject of Appendix~\ref{a:sym}. The reductions to the master integrals
and results for the master spectral functions are given in
Appendix~\ref{a:master}, and the calculation of the master spectral
functions is described in \ref{s:compspec}.  In Appendix~\ref{Dirac} we
show that terms containing a $\gamma^5$ matrix do not contribute to
the Dirac traces of the NLO diagrams.

\textbf{Notation:} 
The signature of the metric is $+ - - - $, 4-vectors are written as
lower-case italics, and bold-face letters refer to 3-vectors.
$ \omega _ n $ are Matsubara frequencies $ \omega _ n = n \pi T
$, with even (odd) integers $ n $ for bosonic (fermionic)
operators. We use the imaginary time formalism where
$ x ^ 0 = - i \tau  $ with real $ \tau $. Matsubara
sums over fermionic frequencies are written as 
$ \sum_{\{k ^ 0\}} F ( k ^ 0 ) \equiv \sum _ { n \rm    \,\,\,odd} F ( i
\omega _ { n } ) $. 

\section{Rates and real-time correlators} 
\label{rates}
\subsection{Matching}\label{matching}

Here we describe how we determine the rates in the effective kinetic
equations (\ref{kineq1}) and (\ref{kineq2}) from the underlying
microscopic quantum field theory by using the theory of
quasi-stationary fluctuations (see, e.g., Ref.~\cite{Landau}, \S 118).
Consider slowly varying quantities $y_{a}$ which va\-nish in thermal
equilibrium.  They satisfy the effective equations of motion
(\ref{eom}).
The thermal fluctuations of $ y _ a   $
observe the same equations, but with an additional
Gaussian white noise term on the right-hand side, which represents the
rapidly fluctuating quantities.  These equations can be used to
compute the real-time correlation function
\begin{equation}
   \mathcal{C}_{  a    b  }(t)=\langle y_{a  }(t)y_{b  }(0)\rangle
\end{equation}
of the fluctuations by
solving these equations and then averaging over  the noise and over 
initial conditions. One obtains
\begin{equation}
  \mathcal{C}_{ a  b }(t)
  =
  \left(e^{-\gamma|t|}\right)_{a  c  }\chi_{cb  }
  \label{ceff} 
  , 
\end{equation}
where the average over initial conditions enters through the 
real and symmetric susceptibilities
\begin{align} 
  \chi_{ab} \equiv \langle y_{a}y_{b}\rangle
  \label{chi} 
  .
\end{align} 
These  susceptibilities have to  be  computed 
in the microscopic theory.
The rate-matrix $ \gamma  $ can be extracted from 
the one-sided Fourier transform 
\begin{equation}
  \mathcal{C}_{ab}^{+}(\omega)
  \equiv \int_{0}^{\infty}dt
  \, e^{i\omega t}\mathcal{C}_{ab}(t)
  \label{c+}
\end{equation}
which is defined for Im $ \omega > 0 $.  For real frequencies $\gamma\ll
\omega \ll\omega_{{\rm UV}},$ where $\omega_{{\rm UV}}$ is the
characteristic frequency of the spectator processes, and for real 
$ \gamma  _ { ab } $ 
one obtains~\cite{bodeker-washout}
\begin{equation}
    \gamma_{ab}
    = 
   \omega^{2}
    \text{ Re } \mathcal{C}_{ac}^{+}(\omega+i \epsilon  )(\chi^{-1})_{cb}
    \quad \mbox{ for } \quad \gamma\ll\omega\ll\omega_{{\rm UV}}
       .
   \label{classic}
\end{equation}
In this regime $   \mathcal{C}_{ab}^{+} $  has to match the one-sided Fourier 
transform of 
the microscopic correlation function
\begin{equation}
   C_{ab}(t)
   \equiv 
   \frac{1}{2}\left\langle \left\{ y_{a}(t),y_{b}(0)\right\} \right\rangle
   .
\end{equation}
The latter can be written as
\begin{equation}
   C_{ab}^{+}(\omega)
   =
   \int
   \frac{d\omega'}{2\pi}\frac{i}
   {\omega-\omega'}
   \left [ 
   \frac{1}{2}+f_{{\rm B}} (\omega')
   \right ] 
   \rho_{ab}(\omega')
   \label{onesided} 
   ,
\end{equation}
with the Bose-Einstein distribution $f_{{\rm B}}(\omega ) \equiv
[ \exp ( \omega  / T ) -1 ] ^{ -1 } $ and the spectral
function (cf.\ Appendix~\ref{a:3pt})
\begin{align}
  \rho_{ab}(\omega) 
  & \equiv 
  \int dte^{i\omega t}
  \left\langle \left[y_{a}(t),y_{b}(0)\right]\right\rangle 
  \label{spec}
  .
\end{align}

So far the discussion is identical to the one
in~\cite{bodeker-washout}.  In~\cite{bodeker-washout}, however, the
spectral function was real. Thus after taking the real part of $ C_{ab}^ +
( \omega + i \epsilon   ) $ only the delta function 
in 
\begin{equation}
\frac{1}{x +i\epsilon}
   =
   -i\pi\delta(x)
   + {\rm P.V.}
   \frac{1}{x}
   \label{deltadisc} 
\end{equation}
contributed to the integral (\ref{onesided}), but not the principal
value.  In this work we consider the spectral function of $X_a$ and
$\delta f_{\vec{k}}$ which have different signs under $CPT$
transformation.  Then the spectral function is imaginary (see Appendix
\ref{a:sym}), and we proceed as follows. Since we are interested in
frequencies much smaller than the temperature $T $, we can approximate
the square bracket in (\ref{onesided}) by $ T/\omega ' $, which gives
\begin{align}
  C ^ + _ { ab } ( \omega  ) 
  =
  -i \frac T \omega  \big [ \Delta    _ { ab } ( \omega  ) - \Delta    _ { ab }
  ( 0 ) \big ]
  .
  \label{quantum} 
\end{align}
Here 
\begin{align} 
 \Delta_{ab}  (\omega  )
 \equiv 
 \int
 \frac{d\omega '}{2\pi} \frac{\rho_{ab}(\omega')}{\omega'-\omega }
 \label{specrepgen}
\end{align} 
is an analytic function off the real axis.  $ \Delta _ { ab } ( \omega
+ i \epsilon ) $ with real $ \omega $ equals the retarded two-point
function.  Matching $ { \cal C } ^ + $ and $ C ^ + $, and using
(\ref{classic}) as well as the fact that $ \Delta _ { ab } ( 0 ) $ is
real we obtain the Kubo-type formula
\begin{equation}
  \gamma_{ab}
  =
  T \omega \, \text{Im} \,\Delta  _{ac}^ {\rm     ret}(\omega)
 \left(\chi^{-1}\right)_{cb}
 \qquad (  \gamma\ll\omega\ll\omega_{{\rm UV}} ) 
 \label{kubo} 
 .
\end{equation}
For real spectral functions this agrees with the 
Kubo-type relation found in~\cite{bodeker-washout}. 

Instead of using (\ref{kubo}) it is a lot more convenient
\cite{bodeker-washout} to compute the retarded correlator $\Pi   _{ab}^{\rm
  ret}(\omega)$ of the time
derivatives $\dot{y}_{a}$, because this way one keeps only the terms which
violate the conservation of $ y _ a   $.
Using $\Pi_{ab}^{\rm ret}(\omega) = \omega  ^ 2
\Delta  _{ ab }^{\rm ret}(\omega)$ we obtain
\begin{equation}
  \gamma_{ab}
  =
  \frac T \omega  \, \text{Im} \, 
  \Pi_{ ac }^{ \rm ret}(\omega)
\left(\chi^{-1}\right)_{cb}
    \qquad (  \gamma\ll\omega\ll\omega_{{\rm UV}} ) 
   \label{Kuborel}
   .
\end{equation}

\subsection{Charge and phase-space density operators}

Among the slow variables we have to consider are charges $ X _ a $
which can be written as
\begin{equation}
  X_{a}
  =\int \! d^3x \, \bar{\lepton }\gamma_{0}T
  _{a} \lepton 
  +\text{contributions from other fermions}
  \label{X} 
  .
\end{equation}
Here $T
_{a}$ is the generator of the corresponding symmetry transformation
acting on the left-handed leptons $ \ell _ i $, where $ i $ is a
family index.  We consider a temperature range in which the charged
lepton Yukawa interactions are either much faster or much slower than
the Hubble rate. In this regime the conservation of the $ X _ a $ is
violated only by the Yukawa-interaction involving the sterile
neutrinos,
\begin{align}
   \mathcal{L}_{\rm int} = 
  {}-   \xbar  {N} \, h 
  \, {\widetilde \varphi}^\dagger \,
   \ell 
   + \mbox{H.c.}
   \label{Lint}
   .
\end{align}
Here $\widetilde \varphi \equiv i \sigma ^ 2 \varphi ^ \ast $ with the
Pauli matrix $ \sigma ^ 2 $ is the isospin conjugate of the Higgs
field $ \varphi $. The Yukawa couplings are written as  a matrix in flavor
space, $ (h ) _ { Ij } = h _ { I j } $.  We describe the sterile neutrinos
by Majorana spinors $ N _ I $.  In (\ref{X}) we have only
written the SU(2) leptons doublets $ \lepton _ i $ explicitly, because
the other Standard Model fermions do not appear in $\mathcal{L} _{{\rm
    int}}$ and do not enter the time derivatives of the  
$  X _ a $. In the Heisenberg picture we obtain
\begin{equation}
  \dot{X}_{a} \left ( x ^ 0 \right ) =i \int d ^ 3 x \left [ 
    Q _ a ( x ) - Q _ a ^\dagger ( x ) \right ], 
  \label{Xdot}
\end{equation}
where
\begin{align}
  Q _ a
  \equiv 
  \xbar {  N } _{I}  \, h _ { I i }  J_{ia}
  \label{Q} 
\end{align} 
with
\begin{align} 
  J_ { i a } 
  \equiv 
   \widetilde{\varphi} ^\dagger( T _  a ) _{ij}\lepton _{j}
  \label{Jia} 
  .
\end{align}

The other slow variables we have to take into account are the
phase-space densities of the sterile neutrinos.  In this work we
consider the hierarchical limit $M_1\ll M_{ I}$ ($ I \neq 1 $), and
temperatures at which only the lightest sterile neutrinos $N_1$ are
present in the plasma.  We assume a homogeneous system, and we neglect
spin asymmetries. Then one can define the phase space density operator
$ f _ { \vec k } $ similarly to~\cite{occupation,asaka}. In the interaction
picture with respect to the Yukawa interaction (\ref{Lint}) the
canonically normalized field can be written as
\begin{equation}
    [ N_{ 1}(x)  ] _ {\rm int } 
   = 
    \sum_ { \vec k, s  } 
    \frac 1 { \sqrt{ 
        2  E _ {  \vec k } V
      }
    }
   \left [e ^{ -ikx } \,   u_{  \vec k s }   \, a _ {  \vec k s} 
     +
      e ^{ ikx } \, v_{  \vec k s}  \, a ^\dagger _ {  \vec k s } 
     \right ] _ { k ^ 0 = E _ {  \vec k  } } 
     \label{NI}
     ,
\end{equation}
where $ E_{\vec k} \equiv ( \vec{k}^{2}+M_{1}^{2} ) ^{ 1/2 } $ .  The
creation and annihilation operators $a ^\dagger _{ \vec{k} s}$ and 
$a _{ \vec{k} s}$ satisfy
\begin{equation}
  \{a_{ \vec{k} s}\,, \,a_{ \vec{k}' s'}^{\dagger}\}
   =
   \delta_{\vec{k}\vec{k}'}
   \delta_{ss'}
   .
\end{equation}
Now we define $ f $ as the spin average
\begin{equation}
  [ f_{\vec{k}} ] _ {\rm int } 
  \equiv 
  \frac 12 
  \sum_{s} a_{ \vec{k} s}^{\dagger}a _{\vec{k}s}
  \label{f} 
  .
\end{equation} 

Switching to the Heisenberg picture, the time derivative of $ f $ can
be obtained from the Heisenberg equation of motion.  For doing perturbation
theory
it is convenient to re-express
the creation and annihilation operators in terms of $ N $ by using
\begin{align} 
   a _ {  \vec k s } 
   = 
   \frac {1 } 
   { \sqrt{  2 E _ {  \vec k } V } } 
   \,
    u ^\dagger _ {  \vec k s } N _ 1 ( 0, \vec k )
   \,, \quad 
   a ^\dagger _ {  \vec k s } 
   = 
   \frac { 1  }
   { \sqrt{  2 E _ {  \vec k } V } }
   \,
   v ^\dagger _ {  \vec k s } N _ 1 ( 0, -\vec k )
  \label{c} 
  ,
\end{align}
where $ N_1 ( t, \vec k) \equiv \int d ^ 3 x e ^{ -i \vec k \vec x }
N_1 ( t, \vec x ) $ are the spatial Fourier-transforms of the
sterile-neutrino field $ N_1 $.  With the definitions
\begin{align}
  R _ {  k } ( t ) 
  \equiv 
  h _ { 1 i } 
  \int \! d ^ 3 x \, d ^ 3 x' 
  e ^{ i \vec k ( \vec x - \vec x' ) }
   \xbar  N _ 1   ( t, \vec x ) \gamma  ^ 0
         &\big (  \slashed{k} + M _ 1 \big )   J _ i(t, \vec x' )
    \label{R} 
\end{align} 
and
\begin{align} 
  J_{i} \equiv \widetilde{\varphi} ^\dagger \lepton _{i}
  \label{Ji} 
\end{align}  
we find\footnote{Note that in {\it real}
  time $  N ^\dagger ( t, - \vec k ) = \big [  N( t, \vec k )
  \big ] ^\dagger  $. \label{Nkreuz}}
\begin{align}
   \dot f _ { \vec k } ( t ) 
   =
   \frac { - i } { 4 V E _ {\vec k }  }
   \Big  \{
      \Big [ R _ { k } ( t ) - R _ { k } ^\dagger ( t )  \Big 
      ] 
 + ( k \to -k ) 
   \Big \} _ { k ^ 0 \to E _ { \vec k }  } 
   \label{fdot}
   .
\end{align}

\section{$CP$ violating rates}\label{general}

\subsection{General considerations}
\label{consider} 

The discussion in Sec.~\ref{rates} applied to all coefficients in
Eqs.~(\ref{kineq1}) and (\ref{kineq2}). Now we will determine the $ CP $
violating ones, $\gamma_{a \vec k }$ and $\gamma_{\vec k a}$, by using
Eq.~\eqref{Kuborel}.  We assume that the slow interaction is the
neutrino Yukawa interaction (\ref{Lint}).  The equal-time correlators
(\ref{chi}) of $ X _ a $ with $ \delta f _ { \vec k } $ vanish due to $
CPT $ invariance, since $ X _ a $ and $ \delta f _ { \vec k } $ are
odd and even under $ CPT $, respectively, and because they commute at
equal times.  Thus the matrix (\ref{chi}) is block-diagonal, and only
the elements
\begin{align}
  \chi_{\vec{k}\vec{k}'}
  & \equiv 
  \langle\delta f_{\vec{k}}\delta f_{\vec{k}'}\rangle 
  ,
  \\
  \chi_{ab}
  &\equiv 
  \langle X_a X_b \rangle
\end{align}
enter (\ref{Kuborel}). At leading order in the Yukawa couplings
$\chi_{\vec{k}\vec{k}'}$ is determined by the free theory which gives
\begin{equation}
  \chi_{\vec{k}\vec{k}'}
  =
  \delta  _ { \vec k \vec k ' } 
  \chi_{\vec k}
\end{equation}
with 
\begin{equation}
  \chi_{ \vec k } 
  = 
  f_{{\rm F}}(E_{\vec k})
  \left [ 1-f_{{\rm F}}(E_{\vec k}) \right ] 
  =
  -Tf_{{\rm F}}'(E_{\vec k})
  \label{chiI} 
  .
\end{equation}
The susceptibility matrix for the charges $\chi_{ab}$ has been
computed in \cite{bodeker-washout,Bodeker:2015zda} up to order $g^2$
in the Standard Model couplings.  Then according to (\ref{Kuborel})
the $CP$ violating rates are given by
\begin{align}
  \gamma_{a\vec{k}}
  =&
  \,
  \frac T { \omega } 
  \text{Im}\,
   \Pi_{a\vec{k}}^ {\rm ret }(\omega)
  \frac 1 {  \chi  _ { \vec k } } 
       ,  \label{asymrate}
       \\ 
  \gamma_{\vec{k}a}
  =&     
  \,
  \frac T { \omega } 
  \text{Im}\,
   \Pi_{\vec{k}b}^ {\rm ret }(\omega)
  \left(\chi^{-1}\right)_{ba} 
  ,
\end{align}
where $ \gamma\ll | \omega | \ll\omega_{{\rm UV}} $.
The  retarded correlators on the right-hand-side  
are given by analytical continuation of the imaginary-time correlators
\begin{align} 
  \Pi_{a \vec{k} }
  (i \omega_{n})
  = &
  \int
  _{0}^{\beta}\!d\tau \, 
  e^{i\omega_{n}\tau}
  \left\langle \dot{X}_{a}(-i \tau)
    \dot{f } _{\vec{k}} (0)\right\rangle
  ,
   \label{correlator}
   \\
 \Pi_{\vec k a }(i\omega_n)
  = &
 \big [ \Pi_{a \vec k }(-i\omega_n)\big ] ^*
 \label{correlator*}
 ,
\end{align} 
where   $ \omega _ n $ is a  bosonic
Matsubara frequency,   and $ \beta \equiv 1/T $.
Now we insert our results 
(\ref{Xdot}) and (\ref{fdot}) for the time derivatives of $ X _ a $
and $ f _ { \vec k } $ to write $ \Pi_{a \vec{k} } $ in terms
of the operators $ Q _ a $ and $ R _ k $ which are defined in Eqs.~(\ref{Q})
and (\ref{R}), respectively. Using
 periodicity in imaginary time  $ x ^ 0 = -i \tau $ we find
\begin{align}
  \left \langle Q _ a ( x ) R ^\dagger _ {k} ( 0 ) \right \rangle 
  = 
  \left \langle Q _ a ^\dagger ( -i \beta  
  - x ^ 0 , \vec x ) R_k ( 0 ) \right \rangle ^ \ast 
  ,
  \label{per} 
\end{align} 
and similarly\footnote{One has to keep in mind that in
  imaginary time $ Q _ a ^\dagger ( x ) = e ^{ i H x ^ 0 } Q _ a ^\dagger ( 0,
  \vec x ) e ^{ -i H x ^ 0 } $ is not the Hermitian conjugate of $ Q _ a
  ( x ) $.} for $ Q _ a ^\dagger ( x ) $. With the help 
of these relations we can simplify $ \Pi_{a \vec{k} } $ such that
it turns into
\begin{align}
  \Pi_{ a \vec k }
  (i \omega_{n})
  &
  = 
    \frac 1 { 2 E _ {\vec k } V } 
  \mbox{Re} 
  \int 
  _ 0 ^ \beta \! d \tau  \int d ^ 3 x \,
    e^{i\omega_{n}\tau} 
  \nonumber \\
  \times 
  &
    \Big \langle 
      \left [ Q _ a( x ) 
        - Q _ a ^\dagger ( x ) \right ] R _ { k } ( 0 ) 
      +  ( k \to -k )  
    \Big \rangle _ { k ^ 0 \to E _ {  \vec k } }
    .
    \label{Pi} 
\end{align}
A non-vanishing value of $X_{a}$ is generated by $ CP $ violating
interactions involving virtual sterile neutrinos $N_{I\neq 1}$ which
first appear in the correlator at order $h^{4}$. Expanding \eqref{Pi}
to order $h^{4}$ we need to expand only to second order in the
interaction (\ref{Lint}) since the operators $ Q _ a $ and $ R _ k $
are  already
linear in $h$.  Using Wick's theorem for the sterile neutrinos, we can
express \eqref{Pi} in terms of free propagators
\begin{align} 
  S_I  ( p )  
  \equiv
  \int  _ 0 ^ \beta \! d \tau  \int \! d ^ 3 x \,
  \,  e ^{ i p x } \left \langle
  N_I(x)\xbar{N}_I(0) \right \rangle
=
  (   \slashed k +M_{I} ) \Delta  _ I ( k ) 
  \label{S} 
  ,
\end{align}
with $ \Delta _ I ( k ) \equiv ( {- k^{2}+M_{I}^{2}} ) ^{ -1 } $, and
four-point correlation functions of the operators (\ref{Jia}),
(\ref{Ji}), which contain only Standard Model fields. These relations of 
the $CP$ violating rates to the four-point function are valid to all orders
in the Standard Model couplings.

\subsection{Hierarchical limit: relation to three-point functions}

In the hierarchical limit $ M _ 1 \ll M _ {  I  \neq 1 } $ one can
integrate out the heavier sterile neutrinos, and work with the
resulting effective theory for $ N _ 1 $ and the Standard Model
fields. We include
only the leading term in $ 1/M _ I $ given by the dimension-5 Weinberg
operator \cite{Weinberg:1979sa}
\begin{align}
  { \cal L } _ 5 
  = 
  \frac 12 c _ { ij } \left ( \overline { \ell _ i ^ c } 
    \widetilde{ \varphi  } ^ \ast 
    \right ) 
  \left ( \widetilde{ \varphi  } ^\dagger \ell _ j \right ) 
  + {\rm H.c. } 
  .
  \label{weinberg} 
\end{align}
Here $ \ell ^ c \equiv \cmatrix \overline \ell ^ \top $ is the charge
conjugate of $ \ell $ with the antisymmetric and unitary
charge-conjugation matrix $ \cmatrix $ which satisfies
\begin{eqnarray}
   \cmatrix ^ { -1 } \gamma  _ \mu  \cmatrix = - \gamma  _ \mu  ^ \top
   \label{cmatrix} 
   .
\end{eqnarray} 
It is convenient to first perform  the contractions described in 
Sec.~\ref{consider}, and
then move on to the effective theory by approximating 
\begin{equation}
  S _ I ( p ) \simeq \frac 1 {M_{I}} \mbox{ for } I \neq1
  \label{SI} 
\end{equation}
and identifying\footnote{Note that the relation (\ref{cij}) is valid
only  for a tree-level matching
of the four-vertex in (\ref{weinberg}) with the corresponding 
4-point function
in the full theory with the interaction
(\ref{Lint}). There are Standard Model
corrections to this relation. In the following we only use the 
effective coupling $c_{ij}$ which should already contain these 
corrections.} 
\begin{align} 
  \sum _ { I \neq 1 } 
  \frac { 
    h _ { I i }  h _ { I j } 
  } { M _ I }
  = 
  c _ { ij } 
  \label{cij} 
  .
\end{align} 
Then the four-point function of Sec.~\ref{consider} 
turns into a three-point function. 
Some of the terms drop out because they  contain the  scalars 
$ \xbar { J _ { \phantom I } } J $ or their Hermitian conjugate,
which vanish because
the operators $ J $ are left-handed.  The remaining  order $h^4$ terms are
\begin{align}
  \int _ 0 ^ \beta  \! d \tau  \int \! d ^ 3 x \,
  &
  e ^{ i \omega  _ n \tau  } 
  \left\langle Q _ a ( x ) R _ {k} \right\rangle 
  = 
  \frac V 2
  \int _ 0 ^ \beta  \! d \tau  \, d \tau  ' \int \! d ^ 3 x \, d ^ 3 x'\,
  T \! \sum _ { n' \rm odd }
  e ^{ i ( \omega  _ n - \omega  _ { n' } ) \tau + i \vec k \vec x  }
  \nonumber \\
  &
  \times h _ { 1 i } h _ { 1j } c ^ \ast _ { lm } 
  \left\langle 
    {\rm T } 
    J _ { j a } ^{  \top } ( x )  \cmatrix S _ 1 ( i \omega  _ { n' } , \vec k )
    \gamma  ^ 0 ( \slashed{ k } + M _ 1 ) J _ i ( 0 ) 
    \xbar  { J _ l }\, \cmatrix \, \xbar  { J _ m } ^ \top ( x ' ) 
    \right\rangle 
    \label{QR} 
\end{align} 
and 
\begin{align}
  \int _ 0 ^ \beta  \! d \tau  \int \! d ^ 3 x \,
  &
  e ^{ i \omega  _ n \tau  } 
  \left\langle Q _ a ^\dagger ( x ) R _{k} \right\rangle 
  = 
  V
  \int _ 0 ^ \beta  \! d \tau  \, d \tau  ' \int \! d ^ 3 x \, d ^ 3 x'\,
  T \! \sum _ { n' \rm odd }
  e ^{ i ( \omega  _ n \tau  - \omega  _ { n' } \tau  ' + \vec k \vec x ') }
  \nonumber \\
  &
  \times 
  h _ { 1 i } h _ { 1 j } c ^ \ast _ { lm } 
  \left\langle 
    {\rm T } 
    \xbar { J _ { l a } } \, \cmatrix  \xbar  {  J _ m } ^ \top ( x ) 
    J _ i ^{\top } ( x ' ) \cmatrix S _ 1 ( i \omega  _ { n' } , \vec k )
    \gamma  ^ 0 ( \slashed{ k } + M _ 1 ) J _ j ( 0) 
    \right\rangle
    \label{QdR} 
    .
\end{align}
Using Eq.~(\ref{S}), and keeping in mind that the $ \ell _ i $ are
left-handed, the products of Dirac matrices in Eqs.~(\ref{QR}) and
(\ref{QdR}) can be simplified as follows,
\begin{align}
  P_{L}S _ 1 \left (  i \omega  _ { n' }, \vec k \right ) 
  \gamma  ^ 0 \left(\slashed k +M_{1}\right) P _ L
  =
  M _ 1 (  i \omega  _ { n' } + k ^ 0 ) 
  \Delta  _ 1 \left (  i \omega  _ { n' }, \vec k \right ) P _ L 
  ,
  \label{L} 
\end{align}
with the left-chiral projector $ P _ L = (1-\gamma  _ 5 ) /2 $. 
With the help of 
\begin{align}
   (  i \omega  _ { n' } + k ^ 0 ) 
  \Delta  _ 1 \left (  i \omega  _ { n' }, \vec k \right )
  = 
  -   \frac 1 {  i \omega  _ { n' } - k ^ 0 } 
  \qquad \mbox{ for  }  \quad k ^ 0 
  = \pm E _    \vec k
  \label{r} 
\end{align} 
one can simplify Eqs.~(\ref{QR}) and (\ref{QdR}) further.  Then we
plug them into Eq.~(\ref{Pi}), thereby replacing $ k ^ 0 \to E _ \vec
k $ after which the variable 
$ k ^ 0 $ no longer appears.  It is then convenient
to rename $ i \omega _ { n' } $ as $ k ^ 0 $, which yields the compact
expression
\begin{align}
  \Pi_{  a \vec k }
   \left (
    i \omega  _ n \right )
    =& \,
      \frac{M _ 1}{2E_{\vec{k}}}
      {\rm Re}
   \biggl \{ 
   T\sum_{\{k ^ 0\}}\frac{1}{k ^ 0-E_{\vec{k}}}
       \biggl (
   \biggl [
   \frac{1}{2}  h_{1i} (hT_{a})_{1j} c ^\dagger _ { lm } 
   \Gamma _ {ijlm}( k - q, q - k)
    \nonumber
    \\
   & {} - h_{1i}h_{1j}  (T_{a}c ^\dagger )_{lm}
      \Gamma _ {ijlm}(k - q, -k)\biggr] 
      - \bigl [ k \to -k \bigr] 
   \biggr ) \biggr \}
   _ { q = (  
        i \omega _ n, \vec  0 ) 
        }
   \label{Xf} 
   ,
\end{align}
with the three-point function
\begin{align}
  \Gamma _ {ijlm}(k_{1},k_{2}) \equiv 
  \int _ 0 ^ \beta  d \tau  _ 1   
  \int _ 0 ^ \beta     d \tau  _ 2  
  &
  \int \! d ^ 3 x _ 1 \int \! d ^ 3 x _ 2 
   e^{i ( k_{1} x_{1} + k _ 2 x _ 2 )  }
   \nonumber \\ &
  \times \left \langle \teuc 
  J_{i}^ \top ( x _ 1 ) \cmatrix ^\dagger  J_{j}(x _ 2 )
  \xbar{J_l}\, \cmatrix \, \xbar {J _ m}^{\top}(0)
   \right \rangle _ 0
  \label{Gamma} 
  .
\end{align}
The time ordering $ \teuc $  is defined in Eq.~(\ref{ordering}), and
the subscript 0 
indicates that the expectation value can now be
evaluated at $ h = 0 $, which is because we are only considering the
leading order in $ h $.

\subsection{Symmetries of the three-point correlator}

To proceed further, we make use of certain symmetries of the
imaginary-time correlation function (\ref{Gamma}).  The two symmetries
\begin{align} 
  \Gamma  _ { ijlm } ( k _ 1, k _ 2 )  
  &=
  \Gamma  _ { jilm } ( k _ 2, k _ 1 )  
  \label{sym1} 
  , \\
  \Gamma  _ { ijlm } ( k _ 1, k _ 2 )  
  &=
  \Gamma  _ { ij ml } ( k _ 1, k _ 2 )  
  \label{sym2} 
\end{align} 
follow from the fact that $ J $ is fermionic and that the matrix $
\cmatrix $ is antisymmetric. They will turn out to be particularly
useful in Sec.~\ref{conti}.  
From now on we neglect  the Standard Model $ CP $ violation.
Then $\Gamma  _ { ijlm } ( k _ 1, k _ 2 )  $ 
is real (see Appendix~\ref{a:sym}). If we further use
that the correlation functions in (\ref{Xf}) do not
depend on the direction of $ \vec k $,  we can write
\begin{align}
  \Pi_{  a  \vec k }
   \left (
    i \omega  _ n \right )
  &
    =
      \frac{i M _ 1}{2E_{\vec{k}}}
   T\sum_{\{k ^ 0\}}\frac{1}{k ^ 0-E_{\vec{k}}}
   \nonumber 
    \\
   \times 
  \biggl \{
  &
    -  \frac{1}{2} 
    \,{\rm Im} \bigl [
       h_{1i} (hT_{a})_{1j} c ^\dagger _ { lm } 
    \bigr ]
   \Gamma _ {ijlm}(-k-q,k+q )
   \nonumber \\
   &
   + {\rm Im} \bigl [
         h_{1i} h_{1j} (T_{a} c ^\dagger )_{lm}
       \bigr ] 
      \Gamma _ {ijlm}(-k-q,k)
      - (  k \to -k ) 
      \biggr \}
      _ { q = (  
        i \omega _ n, \vec  0 ) 
        }
      \label{CP} 
   .
\end{align}
Furthermore, time reversal invariance implies the relation
(\ref{even}). Since $ \Gamma  _ { ijlm
} ( k _ 1, k _ 2 )$ is real,  we have $ \Gamma  _ { ijlm
} ( k _ 1, k _ 2 ) = \Gamma  _ { ijlm
}(
-k _ 1 , -k _ 2 ) $. Thus we find
\begin{align}
  \Pi_{ a \vec k }
   \left (
    i \omega  _ n \right )
  &
    =
      \frac{i M _ 1}{2E_{\vec{k}}}
   T\sum_{\{k ^ 0\}}\frac{1}{k ^ 0-E_{\vec{k}}}
   \nonumber 
    \\
    \times 
    \biggl \{
    &
    -   \frac{1}{2} 
    \,{\rm Im} \bigl [
       h_{1i} (hT_{a})_{1j} c ^\dagger _{lm}
    \bigr ]
   \Gamma _ {ijlm}(- k-q, k + q )
   \nonumber \\
   &
      {}+
{\rm Im} \bigl [
         h_{1i} h_{1j} (T_{a}c ^\dagger )_{lm}
       \bigr ] 
      \Gamma _ {ijlm}(-k- q, k)
      \biggr \}      
_ { q = (  
        i \omega _ n, \vec  0 ) 
        }
      - (  \omega _ n  \to -\omega _ n ) 
      \label{T} 
   .
\end{align}
Now we also  neglect the small Yukawa interactions of the charged
leptons.  Then the remai\-ning Lagrangian is invariant under SU($ N _ {\rm
  fam } $) lepton flavor transformations, where 
$ N _ {\rm fam } =3 $ is the
number of families. Together with the symmetry (\ref{sym2}) this
implies that the three-point correlator (\ref{Gamma}) has the form
\begin{align}
  \Gamma_{ijlm} & 
  =\frac{1}{2}(\delta_{il}\delta_{jm}
  +\delta_{im}\delta_{jl})\Gamma
  ,
  \label{fsym} 
\end{align}
where 
\begin{align}
  \Gamma
  = 
  \frac 1 { N_ {\rm fam } }  \Gamma_{iill}
  \label{contract} 
  .
\end{align} 
Then we find 
\begin{align}
 \Pi_{ a \vec k }(i\omega_{n})&
 =
 M _ 1 
 {\rm Im}\left [ h T _ a c ^\dagger h ^ \top 
 \right ] _ { 11 } 
   {\cal M}_ { \vec{k}} (i\omega_n)
   \label{flavor}
   ,
\end{align}
with
\begin{align}
  {\cal M} _ {\vec{k} }  (i\omega_n)
  = &
  \frac{iT}{4E_{\vec{k}}}
  \sum_{\{k^0\}}
  \left . 
  \frac{
    -\Gamma(-k-q,k+q)
    + 2 \Gamma(-k-q,k)
 }{k^0-E_{\vec{k}}}
 \right | 
 _ { q = (  
   i \omega _ n, \vec  0 ) 
 }
 \nonumber \\ &
 -\left(\omega _ n \rightarrow-\omega _ n\right)
  \label{M}
  .
\end{align}
A nice feature of the formula (\ref{flavor}) is that the effects of
$CP$ violation, described by the imaginary part of the couplings, is
separated from the kinematics, described by the function ${\cal M} _
{\vec{k} } (i\omega_n)$.

\subsection{Matsubara sum and analytic continuation}
\label{conti} 

For the Kubo-type relation (\ref{Kuborel}) we need the retarded
correlator, which can be obtained from the imaginary-time correlator
\eqref{flavor} via analytic continuation.  But first the Matsubara sum
over the  frequency $k_0$ in (\ref{M}) has to be
performed.  This can be done without knowing the three-point
correlators explicitly, by using their spectral representation which
we discuss in more detail in Appendix~\ref{a:3pt}.  The symmetries
(\ref{sym1}) and (\ref{sym2}) imply a relation between the two
spectral functions in (\ref{spectral}), such that one can write
$ \Gamma  $ in terms of a single spectral function, 
\begin{align}
   \Gamma  ( k _ 1, k _ 2 ) 
   =
   &
   \int 
   \frac { d \omega  _ 1 } { 2 \pi  } 
   \frac { d \omega  _ 2 } { 2 \pi  } 
   \frac 1 { k _ 1 ^ 0 + k _ 2 ^ 0 - \omega  _ 1 -\omega  _ 2 }
   \nonumber \\
   & \times 
   \left [ 
     \frac { 
       \rho  ( \omega  _ 1, \vec k _ 1, \omega  _ 2, \vec k _ 2 )      }
        { k _ 1 ^ 0 - \omega  _ 1 } 
     +
     \frac { 
       \rho  ( -\omega  _ 2, - \vec k _ 2, - \omega  _ 1, - \vec k _ 1 )      }
        { k _ 2 ^ 0 - \omega  _ 2 } 
   \right ] 
   \label{Gs} 
   .
\end{align} 
$\rho\equiv \rho_{ABC}$ is the spectral function (\ref{specmom})
for the operators 
\begin{equation}
  A = J^{\alpha}_{i}, \qquad  
  B =  (\cmatrix ^\dagger J_{i})^{\alpha},\qquad  
  C =  \overline{J_l}\cmatrix\overline{J_l}^\top
  \label{ops} 
  ,    
\end{equation}
where $ i $ and the spinor index $\alpha$ are summed over in the
product $AB$. If we insert (\ref{Gs}) in (\ref{M}) we obtain a factor
$ 1/( \omega _ 1 + \omega _ 2 ) $. It does, however, not give rise to
a singularity because it multiplies a function which vanishes for $
\omega _ 1 = - \omega _ 2 $. We can therefore replace it by its
principal value. After that we can re-arrange the integrations until
we obtain
\begin{align}
   {\cal M} _ { \vec k } 
   &
   (i\omega_n)
   =
   \frac{i}{2E_{\vec{k}}}
   \int\frac{d\omega_1}{2\pi}\int\frac{d\omega_2}{2\pi}
   \rho(\omega_1,-\vec{k},\omega_2,\vec{k})
   T\sum_{\{ k_0 \} }\frac{1}{k^0-E_{\vec {k}}}
   \label{Mi}
   \\   &
   \times
   \left(\frac{1}{k^0+\omega_1+i\omega_n}-\frac{1}{k^0+\omega_1}\right)
   \left(\frac{1}{\omega_1+\omega_2+i\omega_n}
     -\pv \frac{1}{\omega_1+\omega_2}\right)
   -
   (\omega_n \rightarrow - \omega_n)
   \nonumber
   .
\end{align}
Now the Matsubara sum is manifestly finite, and we obtain
\begin{align}
  {\cal M} _ { \vec{k} } 
  &
  (i\omega_n)
  =
  \frac{i}{2E_{\vec{k}}} 
  \int\frac{d\omega_1}{2\pi}\int\frac{d\omega_2}{2\pi}
  \rho(\omega_1,-\vec{k},\omega_2,\vec{k})
  \left [ f_{\rm F}(-\omega_1)-f_{\rm F}(E_{\vec k } ) \right ] 
  \label{sum} 
  \\
  \times& 
   \frac{  -i \omega  _ n }
   {
    ( E_{\vec{k}}+\omega_1+i\omega_n )
    ( E_{\vec{k}}+\omega_1 )
    }
\left(\frac{1}{\omega_1+\omega_2+i\omega_n}
  -\pv \frac{1}{\omega_1+\omega_2}\right)
   -(\omega_n\rightarrow -\omega_n)
  \nonumber
  .
\end{align}
In this expression we can analytically continue
$i\omega_n\rightarrow\omega + i \epsilon $ with real $ \omega $.

While the relation  (\ref{asymrate}) is only valid when 
$ | \omega  | \gg 
\gamma   $, where $ \gamma  $ is at most of order $ h ^ 2 $, 
we may now take $ \omega  $ as small as we like 
because $ h $ no longer appears 
in the spectral function $ \rho  $ (cf.\ Eq.~(\ref{Gamma})).
For  small $ \omega
$ we find 
\begin{align}
  {\cal M} _ { \vec{k} } 
  (\omega  + i \epsilon   )
  &
  =
  \frac{i}{E_{\vec{k}}} 
  \int\frac{d\omega_1}{2\pi}\int\frac{d\omega_2}{2\pi}
  \rho(\omega_1,-\vec{k},\omega_2,\vec{k})
  \left [ f_{\rm F}(-\omega_1)-f_{\rm F}(E_{\vec{k}}))\right ] 
\nonumber\\
\times& 
   \frac{  - \omega   }
   {
    ( E_{\vec{k}}+\omega_1+i \epsilon   )
    ( E_{\vec{k}}+\omega_1 )
    }
    ( -i \pi  ) \delta  ( \omega_1+\omega_2 ) 
   + \ord \left (  \omega  ^ 3 \right ) .
   \label{cont} 
\end{align}
We compute the imaginary part of this function, using that $\rho$ is
real valued.  This yields another delta function and leads to the
simple relation
\begin{align} 
  { \rm Im}
 {\cal M} _ { \vec{k} } 
  &
  (\omega  + i \epsilon   )
  =
  -   \frac{\omega  }{4E_{\vec{k}}} 
  \rho(-E _ { \vec k } , -\vec{k}, E _ { \vec k } ,\vec{k})
   f ' _{\rm F}(E_{\vec{k}})
      + \ord ( \omega  ^ 3 ) 
      \label{ImM}
      .
\end{align}
Inserting (\ref{chiI}), (\ref{flavor}) and then (\ref{ImM}) into
(\ref{asymrate}) we finally obtain
\begin{align}
  \gamma_{a\vec{k}} & 
  =
   \frac{\rho(-E_{\vec{k}},-\vec{k},E_{\vec{k}},\vec{k})}{4E_{\vec{k}}}
   M_{1}
   \text{Im}
     \left(hT_{a}c ^\dagger h^ \top \right)_{11}
 \label{mastereff}
 .
\end{align}
This is the {\it  master formula} which relates the asymmetry rate in 
Eq.~(\ref{kineq2}) to the spectral function $ \rho  $ in Eq.~(\ref{Gs}), the
Yukawa couplings $ h _ { 1 i } $, and the dimension-5 
couplings $ c _ { ij } $ in Eq.~(\ref{weinberg}).
Furthermore, from (\ref{correlator*}) and (\ref{sum}) we see that
\begin{equation}
  \Pi_{ \vec k a }(i\omega_n) 
  =
  -\Pi_{  a \vec k }(i\omega_n)
  .
\end{equation}
This implies 
\begin{align}
  \gamma_{\vec{k}a} 
  = 
  -\gamma_{b\vec{k}}  f_{\rm F}(  E_{\vec{k}})
  \left [ 1-f_{\rm F}(E_{\vec{k}}) \right ]  (\chi^{-1})_{ba} 
  \label{master2}
  .
\end{align}
The master formulas (\ref{mastereff}) and (\ref{master2}) have two
important features. First the $ CP $ violation due to the Yukawa
couplings is separated from the kinematic part, which is described by
the spectral function $\rho$. Second the spectral function is exact to
all orders in the Standard Model couplings, except for the very small
charged lepton Yukawa couplings and the Standard Model $ CP $
violation. It can conveniently be computed in finite temperature
perturbation theory.

\section{Lepton asymmetry rate}

\subsection{Leading order at $ T \lsim M _ 1 $}
\label{lepton}

We will now illustrate the use of the master formula (\ref{mastereff})
by computing the leading order asymmetry rate $\gamma_{a\vec{k}}$ in
the regime $ T \lsim M_ 1 $. First we compute the imaginary-time
three-point function, which we analytically continue and then use the
inverse relation (\ref{inv1}) which yields the spectral function $
\rho $ in Eq.~(\ref{mastereff}).

At leading order\footnote{This is the leading order
in the regime $ T \lsim  M _ 1 $. At higher temperature also 
gauge interactions will contribute at leading order like in the sterile neutrino
production rate \cite{anisimov,besak}.}   
the three-point function (\ref{Gamma}) is
given by the diagram
\begin{equation}
  \Gamma^{(0)}(k_1,k_2)=\ToptSEN(\LNb,\Ahh,\Ahh,\Aq,\Aqb,\Lge)
  .
\end{equation}
The solid thick lines represent the operators (\ref{Ji}) which
couple to the sterile neutrinos with outgoing momenta $k_1$ and $k_2$.
The dashed line corresponds to the third operator in (\ref{Gamma})
carrying the outgoing momentum $-k_ 1 -k_2$ which according to
(\ref{mastereff}) will be set to zero in the corresponding spectral
function. The solid arrow-lines are Standard Model leptons and the
dotted lines are Higgses. 

Applying Wick's theorem, computing traces in gauge group and flavor
space and using the property (\ref{cmatrix}) of the charge conjugation
matrix we find
\begin{align}
  \Gamma^{(0)}(k_1,k_2)&
   =
   2\Nweak[ \Nweak+1 ]{\rm Tr} 
   \left(\gamma_{\mu}P_{\rm L}\gamma_{\nu} P_{\rm R} \right)
   I ^  \mu  ( k _ 1 ) I ^ \nu  ( k _ 2 ) 
   \label{LOdiag}
   ,
\end{align}
where $\Nweak=2$ is the dimension of the fundamental representation of the gauge
group SU(2), and
\begin{align} 
  I ^  \mu  ( k ) \equiv \sumint{ \{  p \}   } 
  \frac { p ^ \mu  } { p ^ 2 ( p - k ) ^ 2 } 
  \label{Imu} 
\end{align} 
is a 1-loop sum-integral.  Since the loop integrals are UV divergent
we use dimensional regularization by working with $ d -1 $ spatial
dimensions,
\begin{align} 
  \sumint{ \{ p \} }  
  \equiv 
  T \sum _ { \{ p ^ 0 \}  } 
  \int \frac {  d ^ { d -1 } p } { ( 2 \pi  ) ^{ d -1 } }
  \label{sumint} 
  .
\end{align} 
We have not yet performed the Dirac trace, because it contains
$\gamma^5$ matrices which need a special treatment in dimensional
regularization. We proceed similar to \cite{Laine:2011pq} and use the
definition
\begin{equation}
  \gamma^5
  = -
  \frac{i}{4!}\varepsilon_{\mu\nu\rho\sigma}
  \gamma^{\mu}\gamma^{\nu}\gamma^{\rho}\gamma^{\sigma}
  \label{gam5} 
\end{equation}
of 't Hooft and Veltman \cite{'tHooft:1972fi} and apply the
prescription of \cite{Larin:1993tq} which allows a naively commuting
$\gamma^5$ with $(\gamma^5)^2=1$ in traces with more than one
$\gamma^5$, except in closed fermion loops. Then only traces with one
or no $\gamma^5$ remain. It has been shown in \cite{Laine:2011pq} that
in the trace in (\ref{LOdiag}) all terms with one $\gamma^5$ cancel
exactly due to the total antisymmetry of the Levi-Civita symbol $
\varepsilon_{\mu\nu\rho\sigma} $.  Then the trace becomes
\begin{equation}
  {\rm Tr} \left(\gamma_{\mu}P_{\rm L}\gamma_{\nu} P_{\rm R} \right)
  =
  \frac{1}{2}{\rm Tr} \left(\gamma_{\mu}\gamma_{\nu}\right)
  =
  2\eta_{\mu\nu}
  ,
\end{equation}
and we find the LO order result
\begin{align} 
  \Gamma  
  ( k _ 1, k _ 2 ) 
  =
  - 24 I ^ \mu  ( k _ 1 ) I _ \mu  ( k _ 2 ) 
  \label{LOGam} 
\end{align} 
for the three-point function (\ref{contract}).  The inverse relation
(\ref{inv1}) applied to (\ref{LOGam}) yields the spectral function
\begin{align}
  \rho ^{ ( 0 ) }  ( k _ 1, k _ 2 )
  =
  96\,
   \text{Im}I^\mu(k_1^0+i \epsilon  ,\vec{k} _ 1)   
    \text{Im} I_\mu(k_2^0+i \epsilon  ,\vec{k} _ 2)
  \label{rho}
  ,
\end{align}
\marginpar{537} 
with  real $ k ^ 0 _ 1 $ and $ k ^ 0 _ 2 $. Since $ I _ \mu  ( k ) $
is an even function of $ k $, the spectral function
appearing in the master formula (\ref{mastereff}) becomes
\begin{align}
   \rho ^{ ( 0 ) }  ( - E _ \vec k , -\vec k , E _ \vec k , \vec k )
   =
   - 96 \Big ( {\rm Im} I _ \mu  ( E _ \vec k +i\epsilon , \vec k ) \Big ) ^ 2
   \label{rhoLO} 
   .
\end{align} 
\marginpar{537.1} 

First consider the regime $ T \ll M _ 1 $.  Here the leading
contribution in $ T/M _ 1 $ to the rate $ \gamma _ { a \vec k } $ is
given by the zero-temperature limit of (\ref{mastereff}).  At $ T = 0
$ Lorentz invariance implies
\begin{equation}
  I ^ \mu  ( k ) 
  =
  \frac{k^\mu}{2}I_1(k), 
\end{equation}
with $I_1$ defined in (\ref{I1}). Using Eq.~(\ref{ImI}) we obtain
\begin{equation}
  \rho  ( - E _ \vec k , -\vec k , E _ \vec k , \vec k )
  =  -\frac{24M_1^2}{(16\pi)^{2} } 
  , 
\end{equation}
which together with  the  master formula (\ref{mastereff}) yields
\begin{equation}
  \gamma_{a\vec{k}}^{(0)}
  =
  -\frac{6M_{1}^2}{(16\pi)^{2}E_{\vec{k}}}
  \text{Im}
     \left(hT_{a}c ^\dagger h^ \top \right)_{11}
  .
  \label{LOrate}
\end{equation}
With the tree-level relation (\ref{cij}) for the dimension-5 couplings
this yields
\begin{equation}
  \gamma ^{(0)}
  _{a\vec{k}}
  =
  \varepsilon_a 
  \gamma_{N_1}
  \label{covi} 
  , 
\end{equation}
where
\begin{equation}
  \gamma_{N_{1}}
  =
  \frac{M_1(hh^\dagger)_{11}}{8\pi }
\end{equation}
is the LO $ N _ 1 $ equilibration rate in the non-relativistic
limit $ \vec k ^ 2 \ll M _ 1 ^ 2 $ (see e.g.\ \cite{Bodeker:2013qaa}),
and
\begin{equation}
  \varepsilon_{a}
  =
  -\frac{3}{16\pi}\sum_I\frac{M_{1}}{M_{I}}
  \frac{\text{Im}\left[\left(hT_{a}h^{\dagger}\right)_{1I}
      \left(hh^{\dagger}\right)_{1I}\right]}{(hh^{\dagger})_{11}} 
  .
\end{equation}
Eq.~(\ref{covi}) agrees with the hierarchical limit of the well known
result of Ref.~\cite{Covi:1996wh}.  

Now consider the case $ T \sim M _ 1 $, usually referred to as 
the relativistic regime. After summing over Matsubara frequencies
in Eq.~(\ref{Imu}) one can analytically  continue to real frequencies
which yields
\begin{align}
  {\rm Im} I ^ \mu  ( E _ \vec k +i\epsilon, \vec k ) 
  = 
   \pi  
   &
   \left . \int \! \frac { d ^ 3 p } { ( 2 \pi  ) ^ 3 }
  \delta  ( E _ { \vec k } - | \vec p | - | \vec p - \vec k | )
  \frac { p ^ \mu  } { 4 | \vec p | | \vec p - \vec k | }
  \right | _ { p ^ 0 = | \vec p | } 
  \nonumber \\
  &
  \times \Big  [  1 - f _ { \rm F } ( | \vec p | ) +
  f _ {\rm B } | \vec p - \vec k | ) \Big ] 
  \label{ImuT} 
  .
\end{align} 
\marginpar{537.3} Inserting this in (\ref{rhoLO}), and then in
(\ref{mastereff}) reproduces the LO asymmetry rate for the case $
T \sim M _ 1 $ computed in Ref.~\cite{beneke-density}, where  analytic
results for the integrals
(\ref{ImuT}) were obtained.

\subsection{NLO at zero temperature}
\label{zero}

In this section we calculate the NLO Standard Model corrections to the
three-point spectral function at zero temperature.  This is the leading term
in the low temperature expansion of the 
$ CP $ violating rate $ \gamma _ { a \vec k } $. 
We take into account the U(1) and SU(2)
gauge couplings $g_1$ and $g_2$, the top Yukawa coupling $h_t$ and the
Higgs self-coupling $\lambda$.\footnote{With our convention for the
  hypercharge gauge coupling $ g _ 1 $ the covariant derivative reads
  $ D _ \mu =
  \partial _ \mu + i y _ \alpha g _ 1 B _ \mu + \cdots $, where $ y _
  \varphi = 1/2 $ for the Higgs field and $ B _ \mu $ is the
  hypercharge gauge field. The quartic term in the Higgs potential is
  $ \lambda ( \varphi ^\dagger \varphi ) ^ 2 $.}${}^,$\footnote{ $ O ( g ^ 2
  T^2 ) $ and $ O ( g ^ 2 T^4 ) $ contributions ($ g \in \{ g _ 1, g _
  2, \lambda^{1/2} , y _ t \} $) to the $ CP $ asymmetry in the decays of
  the lightest sterile neutrino have been computed in
  Ref.~\cite{biondini}, but no connection with the kinetic equations
  for leptogenesis was made.}

The gauge interactions give rise to factorizable and 
non-factorizable diagrams. At zero temperature the factorizable diagrams are
\begin{align}
\Gamma_{g,\rm{fac}}(k_1,k_2)
  =&
  \TopGlassesC(\LNb,\Ahh,\Aq,\Ahh,\Ahh,\Ahh,\Aqb,\Agl,\Lge)\nonumber
     +\TopGlassesD(\LNb,\Ahh,\Aqb,\Ahh,\Ahh,\Ahh,\Aq,\Agl,\Lge)\nonumber\\[0.5cm]
     +&\TopGlassesA(\LNb,\Ahh,\Ahh,\Aq,\Aq,\Ahh,\Aqb,\Lgl,\Lge)\nonumber
     +\TopGlassesB(\LNb,\Ahh,\Aq,\Ahh,\Ahh,\Aqb,\Aqb,\Lgl,\Lge)\nonumber\\[0.5cm]
     +&\TopGlassesE(\LNb,\Ahh,\Aqb,\Aq,\Aq,\Aq,\Ahh,\Agl,\Lge)
     +\TopGlassesF(\LNb,\Ahh,\Aq,\Aqb,\Aqb,\Aqb,\Ahh,\Agl,\Lge)
     \label{gfacdiags}
     , 
\end{align}                       
and the non-factorizable ones are\\[1mm]
\begin{align}
  \Gamma_{g,\rm{nfac}}(k_1,k_2)
  =&
  \TopScaryB(\LNb,\Aq,\Aqb,\Ahh,\Ahh,\Agl,\Lhh,\Lhh,\Lge)
                      +\TopScaryB(\LNb,\Ahh,\Ahh,\Aq,\Aqb,\Agl,\Lqb,\Lqb,\Lge)
                      \TopScaryB(\LNb,\Ahh,\Aqb,\Ahh,\Aqb,\Agl,\Lhh,\Lqb,\Lge)
                      +\TopScaryB(\LNb,\Aq,\Ahh,\Aq,\Ahh,\Agl,\Lqb,\Lhh,\Lge)
                      \label{gVdiags}
                      .
\end{align}
The wiggled lines represent electroweak gauge bosons.  Both sets are
independently gauge fixing independent.  The top-quark contributions are
\begin{align}
  \Gamma_{t}(k_1,k_2)
  =&
  \TopGlassesC(\LNb,\Ahh,\Aq,\Ahh,\Aq,\Ahh,\Aqb,\Aq,\Lge)
  +\TopGlassesD(\LNb,\Ahh,\Aqb,\Ahh,\Aq,\Ahh,\Aq,\Aq,\Lge)
  \label{topdiag}
  .
\end{align}
Here the lines in the closed fermion loop represent top-quarks.
At zero temperature the only contribution containing the Higgs self-coupling is 
\begin{equation}
  \Gamma_{\lambda}(k_1,k_2)=\TopScaryA(\LNb,\Aq,\Aqb,\Ahh,\Ahh,\Ahh,\Ahh,\Lge)
  \label{Hidiag}
  .
\end{equation}
Then the  complete NLO correlator is  the sum
\begin{equation}
   \Gamma^{(2)}(k_1,k_2)
   =
   \Gamma_{{ g,\rm fac}}(k_{1},k_{2})
   +
   \Gamma_{{g , \rm  nfac}}(k_{1},k_{2})
   +\Gamma_{t}(k_{1},k_{2})+\Gamma_{\lambda}(k_{1},k_{2})
   . 
   \label{NLO}
\end{equation}
We compute the diagrams with a FORM code~\cite{form} which applies the
following steps.
\begin{enumerate}
 \item Generate the diagrams using Wick's theorem.
 \item Perform traces in flavor and gauge group space.
 \item Insert the expressions for the propagators, with arbitrary  
   gauge fixing parameter $\xi_1$ and $\xi_2$ for $ B $- and $ W $-bosons.

 \item Use the properties of the charge conjugation
   matrix $\cmatrix$. The Feynman-gauge
   results of this step are listed in Appendix~\ref{Dirac} for all
   diagrams.
 \item Perform Dirac traces in naive dimensional regularization. 
      In Appendix~\ref{Dirac} we show that terms with $\gamma^5$
       do not contribute.
 \item Express scalar products in the integrals in terms of inverse
   scalar propagators through the relations
\begin{align}
 p_{i}\cdot p_{j}=&\frac{1}{2}\left(p_i^2+p_j^2-(p_i-p_j)^2\right)
 , \\
 p_{i}\cdot k_1=&\frac{1}{2}\left(p_i^2+k_1^2-(p_i-k_1)^2\right)
 , \\
 p_{i}\cdot k_2=&\frac{1}{2}\left((p_i+k_2)^2-p_i^2-k_2^2\right).
\end{align}
Then all three-point Feynman-integrals appearing in the computation of the
NLO spectral functions have the generic form
\begin{align}
  & I_{a_{1},...,a_{12}}(k_1,k_2)  
  =
  \int_{p_{1},p_{2},p_{3}}
  \frac{1}{p_{1}^{2a_{1}}p_{2}^{2a_{2}}p_{3}^{2a_{3}}
    (p_{1}-k_1)^{2a_{4}}(p_{2}-k_1)^{2a_{5}}(p_{3}-k_1)^{2a_{6}}}
  \nonumber \\
  \times &
  \frac{1}{(p_{1}+k_2)^{2a_{7}}(p_{2}+k_2)^{2a_{8}}(p_{3}+k_2)^{2a_{9}}
    (p_{1}-p_{2})^{2a_{10}}(p_{1}-p_{3})^{2a_{11}}(p_{2}-p_{3})^{2a_{12}}}
\label{masterints}
\end{align}
with integer numbers $a_i$.
 \item Express all scalar integrals in terms of master-integrals.  For
   this step we use the program Reduze \cite{vonManteuffel:2012np}
   which uses the method of integration by parts (IBP)
   \cite{Chetyrkin:1981qh} and the Laporta algorithm
   \cite{Laporta:2001dd} for IBP.  The result of this step in given in
   Appendix~\ref{s:reduction}.
\end{enumerate}

We apply the relation (\ref{invsimple}) to the reduced
three-point functions in Appendix~\ref{s:reduction} and express the
results in terms of master-spectral functions listed in Appendix
\ref{s:masterspecs}.  Then the complete results for the spectral
functions in terms of renormalized couplings are
\begin{align}
  \rho_{g,{\rm nfac}}(-k,k)
  &=
  \frac{3(g_{1}^{2}+g_{2}^{2})M_{1}^{2} \mu ^ { - 4 \varepsilon  } }
  {(16\pi)^{2}8\pi^{2}}
  \left(\frac{3}{\varepsilon}
    +\frac{23}{2}+8\ln(2)+9\ln\frac{\bar{\mu}^{2}}{M_{1}^{2}}\right)
  \label{specnfac}
  ,\\
  \rho_{g,{\rm fac}}(-k,k)
  &=
  -\frac{3(g_{1}^{2}+3g_{2}^{2})M_{1}^{2} \mu ^ { - 4 \varepsilon  }}
  {(16\pi)^{2}4\pi^{2}}
  \left(\frac{3}{\varepsilon}
    +\frac{53}{2}+9\ln\frac{\bar{\mu}^{2}}{M_{1}^{2}}\right)
  ,\\
  \rho_{\lambda}(-k,k)
  &=
  \frac{3\lambda M_{1}^{2}\mu ^ { - 4 \varepsilon  }}
  {\left(16\pi\right)^{2}\pi^{2}}
  \left(\frac{1}{\varepsilon}+\frac{13}{2}
    +3\ln\frac{\bar{\mu}^{2}}{M_{1}^{2}}\right)
  ,\\
  \rho_{t}(-k,k)
  &=
  \frac{3|h_{t}|^{2}M_{1}^{2} \mu ^ { - 4 \varepsilon  }}
  {\left(16\pi\right)^{2}\pi^{2}}
  \left(\frac{3}{\varepsilon}+
    \frac{45}{2}+9\ln\frac{\bar{\mu}^{2}}{M_{1}^{2}}\right)
  \label{spectop}
  ,
\end{align}
with the $\overline{\text {MS}}$ - scale parameter  
$\bar{\mu}^2 \equiv 4\pi \mu^2 e^{-\gamma_{E}}$. Higher orders in 
$\varepsilon \equiv (4-d)/2$ have been neglected.

\begin{figure}[t] 
   \input{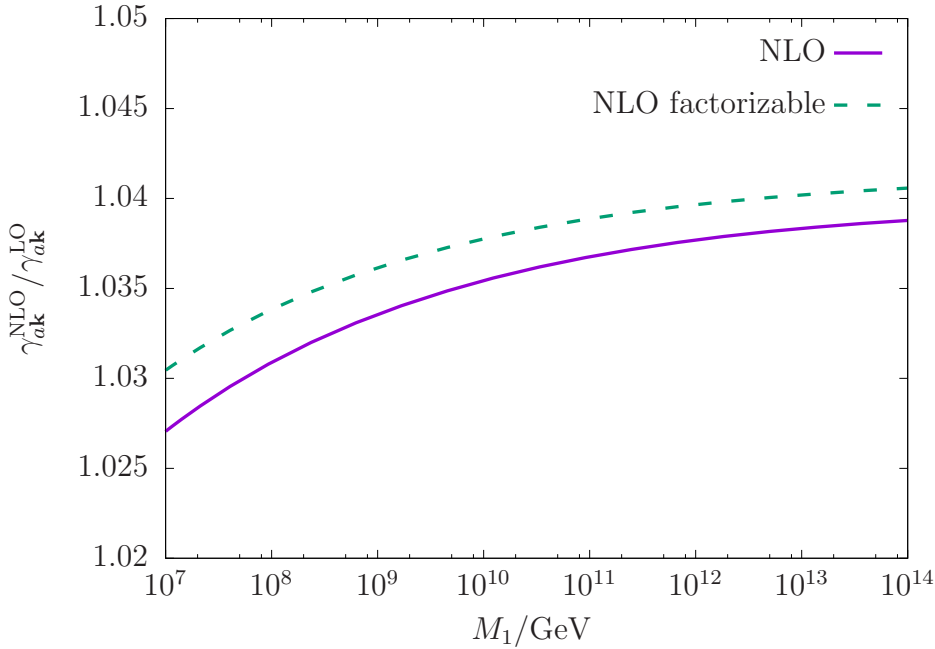}\caption{The relative size of the radiative
     corrections (\ref{res}) to the asymmetry rate in the non-relativistic regime
     $ T \ll M _ 1 $ versus $ M _ 1 $. 
     The radiative corrections are smaller than 4\% over the entire
     mass range relevant to thermal leptogenesis (cf.\
     Ref.~\cite{Blanchet:2008pw}).}\label{plot}
\end{figure}

We renormalize the $ N _ 1 $-Yukawa couplings in the MS-scheme, 
\begin{align}
  h_{1i} 
  = &
  \left (h_{1i} \right )_{\rm R}
  \mu  ^{  \varepsilon  } 
  Z_{h}
  \label{renoYuk1} 
\end{align}
with (see, e.g., Ref.~\cite{grzadkowski}) 
\begin{align}
  Z_{h} 
  & =
  1+\frac{1}{(4\pi)^{2}\varepsilon}
  \left(-\frac{3}{8}(g_{1}^{2}+3g_{2}^{2})+\frac{N_{c}}{2}|h_{t}|^{2}\right)
  \label{Zh}
  ,
\end{align}
and similarly for the dimension-5 couplings, 
\begin{align}
  c _ { ij } 
  =
  \left ( c _ { ij } \right ) _  {\rm R} 
  \mu  ^{ 2 \varepsilon  } Z _ c 
  \label{renoc} 
\end{align} 
with (cf.\ Refs.~\cite{babu,chankowski}) 
\begin{align}
  Z_c
   =
  1+\frac{1}{(4\pi)^{2}\varepsilon}
  \left(
    -\frac{3}{2}g_{2}^{2}
  + 2 \lambda+N_{c}|h_{t}|^{2}\right)
  .\label{Zc}
\end{align} 
We plug the results for the spectral function
(\ref{specnfac})-(\ref{spectop}) into the master formula
(\ref{mastereff}) and express the result in terms of the renormalized
couplings (\ref{renoYuk1}), (\ref{renoc}). Then we obtain the
finite rate
\begin{align}
  \gamma_{a\vec{k}}^{(2)} 
  & =
  \gamma_{a\vec{k}}^{(0)}\Biggl\{1
  +\frac{g_{1}^{2}+3g_{2}^{2}}{(8\pi)^{2}}
  \left(29+6\ln\frac{\bar{\mu}^{2}}{M_{1}^{2}}\right)
  \nonumber\\
  &+\frac{g_{1}^{2}+g_{2}^{2}}{(8\pi)^{2}}
  \left(\frac{1}{2}-8\ln(2)-3\ln\frac{\bar{\mu}^{2}}{M_{1}^{2}}\right)
  \nonumber \\
  & -\frac{|h_{t}|^{2}}{(8\pi)^{2}}
  \left(84+24\ln\frac{\bar{\mu}^{2}}{M_{1}^{2}}\right)
  -\frac{\lambda}{(8\pi)^{2}}
  \left(20+8\ln\frac{\bar{\mu}^{2}}{M_{1}^{2}}\right)\Biggr\}
  \label{res} 
  ,
\end{align}
where $\gamma^{(0)}_{a\vec{k}}$ is the leading-order rate
(\ref{LOrate}) in the effective theory with the interaction
(\ref{weinberg}).  In Fig.~\ref{plot} we show the corrections
(\ref{res}), normalized to the LO result. The renormalization scale is
chosen as $\bar{\mu} = M _ 1$. The corrections are smaller than $4\%$.
The dominant contribution is due to factorizable diagrams, and there
is a cancellation between factorizable and non-factorizable ones.

\section{Summary and outlook}
\label{s:summary}

We have obtained the Kubo-type formula (\ref{Kuborel}), by which one
can relate the $CP$ violating rates in the equations for leptogenesis
(\ref{kineq1}) and (\ref{kineq2}) to finite-temperature real-time
correlation functions. The latter can be systematically computed in
finite-temperature quantum field theory, which allows to include
radiative corrections and determine the theoretical error in
leptogenesis calculations.  For hierarchical sterile-neutrinos masses
$ M _ 1 \ll M _ { I \neq 1 } $ we have expressed the $ CP $ violating
rates in terms of the three-point function (\ref{Gamma}). Using the
spectral representation (\ref{spectral}) we found simple master
formulas (\ref{mastereff}) and (\ref{master2}) relating them to a single
three-point spectral function.  These formulas are valid to leading order in
the sterile-neutrino Yukawa-couplings, and to all orders in the
Standard Model couplings, neglecting the small $CP$ violation and
lepton Yukawa-interactions of the Standard Model.  We applied them to
compute the leading term in the low temperature regime $ T \ll M _ 1 $
of the asymmetry rate $ \gamma _ { a \vec k } $ up to NLO in Standard
Model couplings.  The size of the radiative corrections is smaller
than 4\%.

This work completes the list of Kubo-type relations for the rates in
the kinetic equation for leptogenesis (\ref{kineq1}) and (\ref{kineq2}).
Our low-temperature NLO result is most relevant in the so called
strong washout regime where most of the asymmetry is generated at $ T
< M _ 1 $. There are corrections suppressed by powers of $ T/M _ 1 $,
which should be computable by applying the operator product
expansion~\cite{caron-huot} to Eq.~(\ref{mastereff}). It would also be
interesting, but also a lot more challenging, to compute the NLO in the
relativistic 
regime $ T \sim M _ 1 $. Furthermore, it would be interesting to
compute the LO in the ultra-relativistic regime $ T \gg M _ 1 $.


\section*{Acknowledgments}
We would like to thank M.~Laine for valuable comments and suggestions.

 \appendix

\global\long\def\theequation{\thesection.\arabic{equation}}

\section{Thermal three-point functions and their 
spectral representation}
\label{a:3pt} 

Before we describe the spectral representation for three-point
functions we briefly recall the familiar case of the correlation
function of two operators $A$ and $B$ (for a review see, e.g.,
\cite{Laine:2016hma})
\begin{equation}
  \Delta _{AB}(t_A,t_B) \equiv \langle A(t_A) B(t_B) \rangle
  \label{zwei} 
  .  
\end{equation}
For complex times with
\begin{equation}
  {\rm Im} t_B\geq {\rm Im} t_A\geq -\beta +{\rm Im} t_B
  , 
\end{equation}
it is well defined, 
and its Fourier representation
\begin{equation}
  \Delta_{AB}(t_A,t_B)
  = 
  \int \frac{d\omega_A}{2\pi} 
  \int \frac{d\omega_B}{2\pi} \exp[-i(\omega_A t_A +\omega_B t_B)]
  \Delta_{AB}(\omega  _A, \omega  _B)
  \label{fou} 
\end{equation}
exists.  The cyclicity of the trace in the thermal average
\begin{align} 
  \langle \cdots  \rangle 
  =
  Z ^{ -1 } {\rm tr } [ \exp (  - \beta  H ) \cdots  ] 
  \label{tr} 
\end{align}   
implies 
\begin{align}
  \Delta  _ { AB } ( t _ A, t _ B ) 
  = 
  \Delta  _ { BA } ( t _ B - i \beta , t _ A )
  ,
\end{align} 
and thus
\begin{equation}
  \Delta_{AB}(\omega_A,\omega_B)
  =
  e^{- \beta  \omega_B}\Delta_{BA}(\omega_B,\omega_A) 
  \label{cyc} 
  .
\end{equation}
Translational invariance allows us to write
\begin{equation}
  \Delta_{AB}(\omega_A,\omega_B)
  =
  2\pi\delta(\omega_A+\omega_B)\widehat \Delta_{AB}(\omega_A)
  .
  \label{trans} 
\end{equation}
In this work we pay special attention to the imaginary-time
correlator\footnote{To avoid a proliferation of symbols we use the
  same symbol as in Eq.~(\ref{zwei}). The two can always be distinguished
  by their number of arguments.}\label{proliferation}
\begin{equation}
  \Delta_{AB}(i\omega_n)
  =
  \int_0^\beta d\tau e^{i \omega_n\tau}\left\langle A(-i\tau)B(0)\right\rangle
  \label{Im2pt}
  .
\end{equation}
Using Eq.~(\ref{fou}), (\ref{cyc}), and (\ref{trans}) one
obtains the spectral representation
\begin{equation}
 \Delta_{AB} (i\omega_n)=
 \int
 \frac{d\omega}{2\pi} \frac{\rho_{AB}(\omega)}{\omega-i\omega_n}
 \label{specrep}
\end{equation}
with the spectral function 
\begin{equation}
   \rho_{AB}(\omega )
   \equiv 
   \int 
   dt 
   e^{ i \omega  t } 
   \left \langle\left[A(t),B ( 0 ) \right]\right\rangle
   .
   \label{2ptspec} 
\end{equation}
Here 
\begin{align} 
  [ A, B ] 
  \equiv 
  A B -  ( -1 ) ^{ 
    {\rm deg} A  \,  {\rm deg} B 
  }  B A 
  \label{graded}
  .
\end{align} 
with
\begin{equation}
   {\rm deg}
  A  =  \begin{cases}
0 & \text{if }A \text{ bosonic }\\
1 & \text{if }A \text{ fermionic}
\end{cases}
\end{equation}
denotes the graded commutator of $ A $ and $ B $.  The imaginary time
correlator (\ref{specrep}) can be analytically continued into the
complex frequency plane, $i\omega_n\rightarrow\omega$.  The retarded
and advanced correlators are then given by
\begin{equation}
  \Delta_{AB}^{\rm ret}(\omega)
  = 
  \Delta_{AB}(\omega+i\epsilon),\qquad \Delta_{AB}^{\rm adv}(\omega)
  = \Delta_{AB}(\omega-i\epsilon),
\end{equation}
with real $\omega$.  Then the spectral representation (\ref{specrep})
in combination with the identity (\ref{deltadisc}) 
yields the inverse relation
\begin{equation}
  \rho_{AB}(\omega)
  =
  \frac{1}{i}
  \left [ \Delta^{\rm ret}_{AB}(\omega)-\Delta^{\rm adv}_{AB}(\omega)\right ]
  \label{disc1}
  .
\end{equation}
According to (\ref{specrep}) it is possible to write this inverse relation 
in terms of the imaginary part of 
the retarded correlator as
\begin{equation}
  \rho_{AB}(\omega)
  =
  2{\rm Im}\Delta^{\rm ret}_{AB}(\omega)  \label{disc2}
  ,
\end{equation}
if the spectral function is real-valued.

In the following we derive relations analogous to (\ref{specrep}),
(\ref{2ptspec}) and (\ref{disc1}) for three-point correlators.  To our
knowledge the first spectral representation of three-point functions
at finite temperature has been derived
\cite{Kobes:1990kr,Evans:1990hy} in the real-time formalism for
advanced and retarded correlators. These relations are, however, not
simple integral representations like (\ref{specrep}) and are therefore
not useful for us.  In \cite{Carrington:1996rx,Hou:1997km} three-point
spectral representations have been derived, which are indeed
inte\-gral representations similar to (\ref{specrep}). However, these
references give different spectral-representations for each retarded
and advanced real-time correlator, and none for the imaginary time
correlator.  In \cite{Evans:1990} it has been shown how all six
retarded and advanced three-point functions can be related to the
imaginary-time correlator via analytical continuation. This suggests that
similar to (\ref{specrep}) there is a single spectral representation
for the imaginary-time correlator which also covers the three spectral
representations of \cite{Carrington:1996rx,Hou:1997km}.

We will first derive the spectral representation of the
imaginary-time three-point correlator, using the techniques of
\cite{Evans:1990} and then show that the two independent spectral functions
can be written in terms of (anti-)commutators similar to
(\ref{2ptspec}). Furthermore we show that there are inverse relation
for the spectral functions analogous to (\ref{disc1}).

For operators $ A $, $B $, and $ C $ we consider the  three-point
correlation function 
\begin{align} 
  \Gamma  _ { ABC } ( t _ A, t _ B , t _ C ) \equiv 
   \left \langle A ( t _ A ) B ( t _ B ) C ( t _ C ) \right \rangle
\end{align} 
which is well defined for complex times with 
\begin{align} 
  {\rm Im}\, t _ C  
  \geq
      {\rm Im} \, t_ B\geq{\rm Im}\,t_ A \geq -\beta
  + {\rm Im}\, t _ C
  .
\end{align} 
In this region their Fourier representation
\begin{align}
    \Gamma  _ { ABC } ( t _ A, t _ B , t _ C ) 
   =  
    \int  \frac { d \omega  _ A } { 2 \pi  } 
    \int  \frac { d \omega  _ B } { 2 \pi  } 
    \int   \frac { d \omega  _ C } { 2 \pi  } 
    &
    \,
    \exp \left [  -i ( \omega  _ A t _ A + \omega  _ B t _ B 
      + \omega  _ C t _ C) \right ]  
    \nonumber \\
    & \times  
    { \Gamma }  _ { ABC } ( \omega  _ A, \omega  _ B, \omega  _ C ) 
      \label{fourier} 
\end{align} 
exists.  The cyclicity of the trace in (\ref{tr}) implies 
\begin{align}
  { \Gamma }  _ { ABC } ( \omega  _ A, \omega  _ B, \omega  _ C ) 
    = e ^{ - \beta  \omega  _ C  } \,
    { \Gamma }  _ { CAB } ( \omega  _ C, \omega  _ A, \omega  _ B )
    \label{cyclic} 
    .
\end{align} 
Due to translational invariance in time we can write
\begin{align} 
  { \Gamma }  _ { ABC } ( \omega  _ A, \omega  _ B, \omega  _ C ) 
  = 2 \pi  \delta  ( \omega  _ A + \omega  _ B + \omega  _ C )
  \widehat  \Gamma  _ { ABC } ( \omega  _ A, \omega  _ B ). 
  \label{gamma} 
\end{align} 
We are interested in the Fourier transform of a time ordered
three-point function in imaginary time,\footnote{Cf.\ the footnote on
page \pageref{proliferation}.}
\begin{align}   
  \Gamma  _ { ABC } ( i \omega  _ n , i \omega  _ { n' }  ) 
  \equiv 
  \int _ 0 ^ \beta  d \tau  
  \int _ 0 ^ \beta  d \tau  '
  \,
  \exp  ( i \omega  _ n \tau   + i \omega  _ { n' }  \tau  ' ) 
    \left \langle  {\rm T }
      A ( -i \tau    ) B ( -i \tau  ' ) C ( 0 ) \right \rangle
    \label{eucl3pt} 
    ,
\end{align} 
where the time ordering T is defined as 
\begin{align} 
  {\rm T }
      A ( -i \tau    ) B ( -i \tau  ' )
  \equiv & \,\,
        \theta  ( \tau   - \tau  ' ) 
      A ( -i \tau    ) B ( -i \tau '   ) 
      \nonumber \\ {}  & 
   + ( -1 ) ^{ {\rm deg}  A \, {\rm deg}  B  }
   \theta  ( \tau  ' - \tau  ) 
       B ( -i \tau   ' ) A ( -i \tau    ) 
       \label{ordering} 
    .
\end{align} 
Following \cite{Evans:1990}, we use the Fourier representation
(\ref{fourier}) of the correlators on the right-hand side to perform 
the $\tau$ and $\tau '$ integrals. This gives an integral
representation of the imaginary-time correlator, containing the real-time
correlation functions $\gamma_{ABC}$ and $\gamma_{BAC}$ and
temperature-dependent exponential functions.  Unlike
\cite{Evans:1990}, we rearrange the pre-factors of the real-time
correlators using the cyclicity property (\ref{cyclic}) which allows
us to cancel all exponentials. This yields the simple spectral
representation
\begin{align}
   \Gamma _ { ABC }  ( i \omega _ n  , i \omega _ { n' }  ) 
   = 
    \int \frac { d \omega _ 1  } { 2 \pi  } 
    &
    \int \frac { d \omega  _ 2 } { 2 \pi  } 
    \,
   \frac 1 { i \omega  _ n + i \omega  _ { n' }    -\omega _ 1  - \omega _ 2  }
   \nonumber \\
   & \,
   \times \left [ \frac { \rho  _ { ABC } ( \omega _ 1 , \omega  _ 2 ) }
     { i \omega  _ n - \omega _ 1  } 
      + ( -1 ) ^{ {\rm deg}  A \, {\rm deg}  B  }
     \,
     \frac { \rho  _ { BAC } ( \omega _ 2 , \omega _ 1 ) }
     { i \omega  _ { n' }  - \omega _ 2 } 
     \right ] 
     \label{spectral} 
     . 
\end{align} 
It contains two spectral functions
\begin{align}
    \rho  _ { ABC } ( \omega  , \omega  ' )
    \equiv 
    \int \! d t \int \! d t ' 
    \,
    \exp (  i \omega   t  + i\omega ' t ' ) 
    \left \langle
      \Big [ A ( t  ) , \big [ B ( t ' ) , C ( 0 ) \big ] 
        \Big  ] 
        \right \rangle 
        \label{specmom} 
\end{align} 
which contain the graded commutators (\ref{graded}).  Like in the case
of two-point functions, the three-point spectral functions
(\ref{specmom}) can be computed from the imaginary time correlator via
analytic continuation.  This yields six different retarded functions
which in the notation of \cite{Evans:1990} read
\begin{align}
   &R_1(\omega_A,\omega_B)
    =
   \Gamma _{ABC}
   (\omega_A+2i\epsilon, \omega_B-i\epsilon)
   \label{R1}
   \\
   &R_2(\omega_A,\omega_B)
    =
   \Gamma  _{ABC} (\omega_A-i\epsilon, \omega_B+2i\epsilon)
   &
   \label{R2}
   \\
   &R_3(\omega_A,\omega_B)
   =
   \Gamma _{ABC}  (\omega_A-i\epsilon, \omega_B-i\epsilon)
   &
   \label{R3}
   \\
   &
   \xoverline { R } _ i (\omega_A,\omega_B)
   = 
   R _ i (\omega_A,\omega_B) \Big | _ { \epsilon  \to -\epsilon  } 
   &
   ( i= 1, \ldots  ,3 ) 
   .   \label{Rb} 
\end{align} 
In contrast to the integral representation in \cite{Evans:1990} our
spectral representation (\ref{spectral}) provides simple inverse
relations. We obtain them by combining (\ref{spectral}) and
(\ref{deltadisc}), and we easily find
\begin{align}
  \rho_{ABC} 
  =&
  R_2 +\xoverline {R } _ 2 - R_3 - \overline {R}_3
  ,
  \label{inv1}\\
  \rho_{BAC}
  =& (-1)^{{\rm deg} A \, {\rm deg} B}
  (
  R_1 +  \xoverline {R}_1
  - R_3 - \xoverline {R}_3 )
  .
  \label{inv2} 
\end{align}
If the spectral function is real, these relations can be further
simplified to
\begin{align}
   \rho_{ABC} 
   & =
   2{\rm Re}(  R_2 -R_3  ) 
   \label{invsimple}
   \\
   \rho_{BAC}
   & =
   (-1)^{{\rm deg} A \,{\rm deg} B}
   2\text{Re}(R_1-R_3) 
   .
\end{align}

Let us summarize the results of this section. We have shown that the
imaginary-time three-point correlator (\ref{eucl3pt}) can be expressed in
terms of a single spectral representation (\ref{spectral}), containing
two independent spectral functions.  The retarded and advanced
correlators are related to the imaginary-time correlator by
(\ref{R1})-(\ref{Rb}) and yield the inverse relations (\ref{inv1}) and
(\ref{inv2}).

\section{Implications of  discrete symmetries for spectral functions}
\label{a:sym}

First consider two-point spectral functions of Hermitian
operators $ X $ which satisfy
\begin{equation}
  \cc\pa\tr X(t) (\cc\pa\tr)^{-1}=\vartheta_X  X(-t) 
  \label{CPTtrafo}
\end{equation}
with a phase factor  $ \vartheta _ X$. $ CPT $ invariance implies
that 
\begin{align} 
  \langle { \cal O } \rangle = \left \langle \cc\pa\tr { \cal O }
    (\cc\pa\tr)^{-1} \right \rangle ^ \ast 
\end{align}
for the thermal
expectation value of any ope\-rator $ \cal O $, and in our case
\begin{equation}
  \rho_{AB} (\omega)
  =
  \vartheta_A ^ \ast \vartheta _B ^ \ast  \big [ \rho_{AB}(\omega) \big ]  ^*
  \label{rhorel} 
  .
\end{equation}
Thus the two-point spectral functions are real or
imaginary if $ \vartheta _ A \vartheta _ B = 1 $ or $ \vartheta _ A
\vartheta _ B = -1 $, respectively.\footnote{In \cite{Kadanoff} this
  was shown under the much stronger assumption of time-reversal
  invariance.}  For  the case $A=X_a$ and $B=\delta
f_{\vec{k}}$, defined through (\ref{X}) and (\ref{f}), the spectral
function is imaginary.

Now we turn to   three-point correlators and their
spectral functions. Here we assume $ CP $  and thus $ T $ invariance.
We are interested in field operators, not necessarily Hermitian, which satisfy
\begin{align}
  \cc\pa   \big[A(t _ 1, \vec x _ 1 ),
    \left[B(t _ 2, \vec x _ 2),C(0)\right]\big] (\cc\pa)^{-1} 
  & =
  \vartheta_{CP}
  \big[A( t _ 1, -\vec x _ 1 ),\left[B(t_2, -\vec x _ 2 ),C(0)
      \right]\big] ^\dagger 
  \label{cpabc} 
  ,
  \\
  \tr  \big[A(t _ 1, \vec x _ 1 ),
    \left[B(t _ 2, \vec x _ 2), C(0)\right]\big]  \tr ^{-1} 
  & =
  \vartheta_{T} \big [ A(-t _ 1, \vec x _ 1),
    \left[B(-t_ 2, \vec x _ 2 ),C(0)\right]\big]
  ,
\end{align}
with $\vartheta_{T}$ and $\vartheta_{CP}$ being $\pm1$.  Since the
operators $ A $, $ B $ depend on spatial coordinates $ \vec x _ 1 $
and $ \vec x _ 2 $, the corresponding spectral function (cf.\
Eq.~(\ref{Gs})) now also depends on the conjugate variables $ \vec k _
1 $ and $ \vec k _ 2 $.  $ CP $ invariance implies
\begin{align}
   \rho_{ABC}(\omega  _ 1, \vec k _ 1, \omega  _ 2, \vec  k _ 2 )  
  & =
  \vartheta_{CP}\big [ 
  \rho_{ABC}(- \omega   _ 1 , \vec k _ 1, - \omega   _ 2, \vec k _ 2) 
  \big ] ^{*}
  \label{cprho}
  , 
\end{align} 
and time reversal invariance gives
\begin{align} 
   \rho_{ABC}(\omega_1, \vec k _ 1,\omega_ 2, \vec k _ 2) 
  & =
  \vartheta_{T}
  \big [ 
    \rho_{ABC}(\omega _ 1, - \vec k _ 1, \omega_ 2, -\vec k _ 2)
    \big ] ^ \ast
  \label{trho} 
  .
\end{align}
Assuming that $ [ A, [ B, C ] ] $ is a scalar, 
rotational invariance allows us to rewrite this as
\begin{align} 
   \rho_{ABC}(\omega_1, \vec k _ 1,\omega_ 2, \vec k _ 2) 
  & =
  \vartheta_{T}
  \big [ 
    \rho_{ABC}(\omega _ 1,  \vec k _ 1, \omega_ 2, \vec k _ 2)
    \big ] ^ \ast
  \label{rrho} 
  .
\end{align}
Thus, depending on the sign of $ \vartheta _ T $, the spectral function
is either real or imaginary.  Combining Eqs.~(\ref{cprho}) and 
(\ref{trho})  we obtain
\begin{align} 
   \rho_{ABC}(\omega_1, \vec k _ 1,\omega_ 2, \vec k _ 2) 
  & = 
  \vartheta_{CP}\vartheta_{T} 
  \rho_{ABC}(-\omega_1, \vec k _ 1,-\omega_ 2, \vec k _ 2) 
  \label{cptrho}
  .
\end{align} 
Therefore the spectral function is either even or odd under 
$(\omega_1,\omega_2)\rightarrow(-\omega_1,-\omega_2).$ 

Now we specialize to the three-point function (\ref{Gamma}), which enters 
the asymmetry rate. 
Here we can use 
\begin{align} 
  \cc\pa J _ i(t,\vec{x})( \cc\pa ) ^{-1}  
  = & 
  \eta _ { CP }  \cmatrix \big [   J _ i (t,-\vec{x}) \big ]   ^ \ast
  \label{cpj} 
  \\
  \tr J _ i(t,\vec{x})\tr ^ {-1}  
  = & 
  \eta  _ T   \gamma_{1}\gamma_{3}J _ i(-t,\vec{x})
  ,
  \label{tj} 
\end{align} 
where $ \cmatrix $ is the charge conjugation matrix, $ 
\gamma  _ 1 $, $ \gamma  _ 3 $ are Dirac matrices, and
$\eta _ { CP } $ and $ \eta _ T $ are  phase factors. Eqs.~(\ref{cpj}),
(\ref{tj}) imply 
$\vartheta_{CP}= \vartheta_{T}=1 $ for the phase factors in
Eq.~(\ref{cptrho}). Thus the
corresponding spectral function $\rho_{ABC}$ and $\rho_{BAC}$ are real
and even under $(k_{1},k_{2})\rightarrow(-k_{1},-k_{2}).$ Using the
spectral representation (\ref{spectral}) and Eq.~(\ref{trho}) one
finds
\begin{align}
  [ \Gamma   _ { ijlm}(k_{1},k_{2}) ] ^ \ast 
  & =
  \Gamma  _ { ijlm} (-k_{1},-k_{2})
  \label{even} 
  .
\end{align}
\marginpar{454.1} 
Furthermore, in imaginary time Eq.~(\ref{cprho}) implies that
\begin{align}
  \big [  \Gamma  _ { ijlm}(k_{1},k_{2}) \big ]  ^{*}
  &  = 
  \Gamma  _ { ijlm } (k_{1},k_{2})
  \label{real}
  .
\end{align}

\section{Master integrals and master spectral functions} 
\label{a:master} 

\subsection{Results of the reduction to master integrals}
\label{s:reduction}

We use the program Reduze \cite{vonManteuffel:2012np} to obtain the
following gauge-parameter independent contributions
to  the three-point
correlator (\ref{contract})  in terms of master integrals, defined in
(\ref{masterints}).  For the factorizable  diagrams in Eq.~(\ref{gfacdiags}) 
we find
\begin{align}
  \Gamma_{g,{\rm fac}}(k_1,k_2)
  = & 2\Nweak(\Nweak+1)(y^2_{\tilde{\varphi}}g_1^2+C_2(r)g_2^2)
  \nonumber \\
  \times & 
  \biggl(\frac{(d-2)(-4-d+d^{2})k_2\cdot k_1}
  {(-4+d)^{2}k_2^{2}}I_{011001100100}
  \nonumber \\
  + & 
  \frac{(d-2)(-4-d+d^{2})k_2\cdot k_1}{(-4+d)^{2}k_1^{2}}I_{101010001100}
  \nonumber \\
  - & 
  \frac{(d-2)k_2\cdot k_1}{(d-4)}I_{111011100000}
  -  \frac{(d-2)k_2\cdot k_1}{(d-4)}I_{111001110000}\biggr)
  ,
  \label{RedFac}
\end{align}
and for the non-factorizable ones in Eq.~(\ref{gVdiags}) 
\begin{align}
  \Gamma_{g,{\rm nfac }}(k_1,k_2) 
  & =
  -2\Nweak(y_{\tilde{\varphi}}y_{\ell}(\Nweak+1)g_1^2+C_2(r)g_2^2)
  \nonumber \\
  \times & 
  \biggl(
  \frac{-(d-2)(2d-5)(-20+79d-48d^{2}+8d^{3})}
  {(d-3)^{2}(3d-10)(3d-8)k_2^{2}}
  I_{001000010110}
  \nonumber \\
  {}+ 
  & 
  \frac{4(d-2)(2d-5)(2d-3)}
  {(d-4)(3d-8)k_2^{2}}
  I_{000001010110}
  \nonumber \\
  {} + 
  & 
  \frac{(d-2)(2(9-9d+2d^{2})(k_1+k_2)^{2}+(-25+23d-5d^{2})k_2^{2})}
  {(d-3)(3d-8)k_2^{2}}
  I_{001001010110}
  \nonumber \\
  + & 
  \frac{(42560-78192d+58256d^{2}-22318d^{3}+4561d^{4}-456d^{5}+16d^{6})}
  {(d-4)^{2}(d-3)^{2}(3d-10)(3d-8)k_1^{2}}\nonumber\\
  &\times (2d-5) I_{010001000110}\nonumber \\
  + & \frac{-2320+2900d-1168d^{2}+93d^{3}+41d^{4}-7d^{5}}
  {(d-4)(d-3)(3d-10)(3d-8)}I_{010001010110}\nonumber \\
  + & 
  \frac{2(-60+55d-15d^2+d^3)k_1^{2}-2(d-2)^{2}(4d-13)k_1\cdot k_2}
  {(3d-8)^{2}}I_{011001010110}\nonumber \\
  + & \frac{(d-2)^{2}}{(d-3)}I_{011001100100}
  +  \frac{8(-60+55d-15d^{2}+d^{3})k_1^{2}k_2^{2}}{(3d-10)(3d-8)^{2}}
  I_{021001010110}\nonumber \\
  + & \frac{-4(d-2)}{d-4}I_{100001010110}
  +  \frac{(d-2)k_1^{2}}{(d-3)}I_{101001010110}
  +  \frac{(d-2)k_2^{2}}{(d-3)}I_{110001010110}\nonumber \\
  + & 
  \frac{(d-2)(2d-3)k_1^{2}(k_1+k_2)^{2}}{(d-3)(3d-8)k_2^{2}}
  I_{001002010110}
  \nonumber \\
  + & \frac{(12-5d)(k_1+k_2)^{2}}{(d-3)(3d-8)}I_{010002010110}
  +  \frac{(d-2)^{2}}{(d-3)}I_{101010001100}\biggr)
   ,
   \label{RedNfac}
\end{align} 
for the top-quark contribution (\ref{topdiag}) 
\begin{align}
\Gamma_{t}(k_1,k_2) & =2\Nweak(\Nweak+1)N_c|h_t|^2\nonumber \\
\times & \biggl(\frac{(d-2)k_2\cdot k_1}{(d-4)k_1^{2}}I_{101010001100}
+  \frac{(d-2)k_2\cdot k_1}{(d-4)k_2^{2}}I_{011001100100}\biggr)
   .
   \label{RedTop}
\end{align}
and 
for the Higgs-contribution in (\ref{Hidiag}) 
\begin{align}
  \Gamma_{\lambda}(k_1,k_2)
  = & 
  -4\Nweak(\Nweak+1)\lambda
  \nonumber \\
  \times &
  \biggl( \frac{(-2+d)(-5+2d)(-20+7d)}{(-3+d)(-10+3d)(-8+3d)k_2^{2}}
  I_{001000010110}
  \nonumber \\
  + & 
  \frac{-4(-2+d)(-5+2d)}{(-4+d)(-8+3d)k_2^{2}}I_{000001010110}
  +  \frac{(-2+d)}{(-8+3d)}I_{001001010110}
  \nonumber \\
  + & 
  \frac{-(-5+2d)(-1040+1064d-362d^{2}+41d^{3})}
  {(-4+d)(-3+d)(-10+3d)(-8+3d)k_1^{2}}
  I_{010001000110}
  \nonumber \\
  - & 
  \frac{(100-72d+13d^{2})}{(-10+3d)(-8+3d)}I_{010001010110}
  -  \frac{(-4+d)(k_1+k_2)^{2}}{(-3+d)(-8+3d)}I_{010002010110}
  \nonumber \\
  +    & 
  \frac{-2((-4+d)(-5+2d)k_1^{2}+(-2+d)^{2}k_1\cdot k_2)}{(-8+3d)^{2}}
  I_{011001010110}
  \nonumber \\
  + & \frac{-8(-4+d)(-5+2d)k_1^{2}k_2^{2}}{(-10+3d)(-8+3d)^{2}}
  I_{021001010110}
  \nonumber \\
  - & \frac{(-2+d)k_1^{2}(k_1+k_2)^{2}}{(-3+d)(-8+3d)k_2^{2}}
  I_{001002010110}
  \biggr)
  .
  \label{RedHi}
\end{align}
Here $ \Nweak = 2 $ is the dimension of the fundamental representation
of SU(2).

\subsection{Results for master spectral functions}
\label{s:masterspecs}

The only master integrals in Appendix \ref{s:reduction}
which contribute to the spectral functions (\ref{specnfac}) -
(\ref{spectop}) are
\begin{align}
  I_{{\rm BB}}(k_1,k_2) & \equiv I_{011001010110}(k_1,k_2),\\
  I_{{\rm BBdot}}(k_1,k_2) & \equiv I_{021001010110}(k_1,k_2),\\
  I_{{\rm LR}}(k_1,k_2) & \equiv I_{100001010110}(k_1,k_2),\\
  I_{{\rm 2L}}(k_1,k_2) & \equiv I_{110001010110}(k_1,k_2),\\
  I_{2{\rm R}}(k_1,k_2) & \equiv I_{101001010110}(k_1,k_2),\\
  I_{{\rm fac3L}}(k_1,k_2) & \equiv I_{101010001100}(k_1,k_2),\\
  I_{{\rm fac3R}}(k_1,k_2) & \equiv I_{011001100100}(k_1,k_2),\\
  I_{{\rm fac4L}}(k_1,k_2) & \equiv I_{111011100000}(k_1,k_2),\\
  I_{{\rm fac4R}}(k_1,k_2) & \equiv I_{111001110000}(k_1,k_2),
\end{align}
with $I_{a1,...,a12}(k_1,k_2)$ defined in (\ref{masterints}).  For all
other master integrals we find that either the corresponding spectral
function $\rho(-k,k)$ vanishes, or the integrals are multiplied by
$(k_1+k_2)^2$ which is put to  zero at the end of the
calculation. We have checked that such spectral functions do not have
a $1/(k_1+k_2)^2$ pole, which could cancel the factor $(k_1+k_2)^2$.
Following the steps of Appendix~\ref{s:compspec} we find for the
master spectral functions expanded in $\varepsilon=(4-d)/2$
\begin{align}
  \rho_{{\rm BB}}(-k,k) 
  & 
  =
  -\frac { \mu  ^{ -6 \varepsilon  } }{(16\pi)^{2}4\pi^{2}}
  \left [ 
    \frac{1}{\varepsilon}+7+3\ln\left(\frac{\bar{\mu}^{2}}{k^{2}}\right)
    \right ],\label{rhoBB}
  \\
  \rho_{{\rm BBdot}}(-k,k) 
  & = 
  \frac{\mu  ^{ -6 \varepsilon  }}{(16\pi)^{2}4\pi^{2}k^{2}}
  \left [ 
    \frac{1}{\varepsilon}+4+3\ln\left(\frac{\bar{\mu}^{2}}{k^{2}}\right)
    \right ], \label{rhoBBdot}
  \\
  \rho_{{\rm LR}}(-k,k) 
  & =
  \frac{k^{2} \mu  ^{ -6 \varepsilon  } }{(16\pi)^{2}8\pi^{2}}
  \left \{  
    1+\varepsilon\left [ 
      10+3\ln\left(\frac{\bar{\mu}^{2}}{k^{2}}\right)
      \right ] 
    \right \},\label{rhoLR}
    \\
    \rho_{2L}(-k,k) & 
    = 
    -\frac{1+\ln(2)}{(16\pi)^{2}4\pi^{2}}
    \mu  ^{ -6 \varepsilon  }
    \label{rho2L},\\
    \rho_{2R}(-k,k) 
    & =
    -\frac{1+\ln(2)}{(16\pi)^{2}4\pi^{2}} \mu  ^{ -6 \varepsilon  }
    \label{rho2R},\\
    \rho_{{\rm fac3L}}(-k,k) 
    & =
    \frac{k^{2} \mu  ^{ -6 \varepsilon  }}{(16\pi)^{2}8\pi^{2}}
     \left\{1+\varepsilon
     \left[
     \frac{17}{2}+3\ln\left(\frac{\bar{\mu}^{2}}{k^{2}}\right)
     \right]
     \right\}
     ,
     \label{rho3L}
     \\
     \rho_{{\rm fac3R}}(-k,k) 
     & =
     \frac{k^{2} \mu  ^{ -6 \varepsilon  }}{(16\pi)^{2}8\pi^{2}}
     \left\{1+\varepsilon
     \left[
     \frac{17}{2}+3\ln\left(\frac{\bar{\mu}^{2}}{k^{2}}\right)
     \right]
     \right\},\label{rho3R}\\
     \rho_{{\rm fac4L}}(-k,k) 
     & =
     -\frac{\mu  ^{ -6 \varepsilon  }}{(16\pi)^{2}2\pi^{2}}
     \left[
     \frac{1}{\varepsilon}+6+3\ln\left(\frac{\bar{\mu}^{2}}{k^{2}}\right)
     \right]
     ,\label{rho4L}\\
     \rho_{{\rm fac4R}}(-k,k) 
     & =
     -\frac{\mu  ^{ -6 \varepsilon  }}{(16\pi)^{2}2\pi^{2}}
     \left[
     \frac{1}{\varepsilon}+6+3\ln\left(\frac{\bar{\mu}^{2}}{k^{2}}\right)
     \right]
     .
     \label{rho4R}
\end{align}
                             
\section{Computation of master spectral functions}
\label{s:compspec}

In this section we explain the method which we used to compute
the master three-point spectral functions in Appendix~\ref{s:masterspecs}.

\subsection{Factorizable integrals}

If the master integrals can be written as a product of two-point
integrals $I_a(k_1)$ and $I_b(k_2)$ as
\begin{equation}
  \Gamma(k_1,k_2)=I_a(k_1)I_b(k_2),
\end{equation}
one can simplify the inverse relation (\ref{invsimple}) to
\begin{align}
  \rho(k_1,k_2) & =\rho_{a}(k_1)\rho_{b}(k_2)
  ,\label{invFac}
\end{align}
where 
\begin{equation}
  \rho_{a}(k)=2 {\rm Im} I_a(k_0+i\epsilon,\vec{k})
\end{equation}
is the two-point spectral function of the integral $I_a$. 
In this work we have to deal with the factorizable integrals
\begin{align}
I_{{\rm fac3L}}(k_1,k_2) & =I_3(k_1)I_1(k_2)\\
I_{{\rm fac3R}}(k_1,k_2) & =I_1(k_1)I_3(k_2)\\
I_{{\rm fac4L}}(k_1,k_2) & =I_2(k_1)I_1(k_2)\\
I_{{\rm fac4R}}(k_1,k_2) & =I_1(k_1)I_2(k_2), 
\end{align}
where
\begin{align}
  I_1(k)
  &=
  \int\frac{d^{d}p}{(2\pi)^{d}}\frac{1}{p^{2}(p-k)^{2}}   
  \label{I1}  \\
  I_2(k)
  &=
  \int\frac{d^{d}p_1}{(2\pi)^{d}}
  \frac{d^{d}p_2}{(2\pi)^{d}}\frac{1}{p_1^{2}(p_1-k)^{2}p_2^{2}(p_2-k)^{2}}
  \\
  I_3(k)
  &=
  \int\frac{d^{d}p_1}{(2\pi)^{d}}\frac{d^{d}p_2}{(2\pi)^{d}}
  \frac{1}{p_1^{2}(p_1-p_2)^2(p_2-k)^{2}}
  .
\end{align} 
The imaginary parts of their analytic continuation to real $ k ^ 0 $ 
have been computed in  \cite{Laine:2011pq} and read
for $ k ^ 2 > 0$
\begin{align}
  {\rm Im}I_1(k ^ 0 +i \epsilon  ,\vec{k})
  &
  =
  \frac{{\rm sgn}(k^0)}{16\pi}
  \mu  ^{ -2 \varepsilon  } 
  \left[1+\varepsilon\left(\ln\frac{\bar{\mu}^2}{k^2}+2\right)\right]
  +\ord (\varepsilon^2)
  \label{ImI}
  \\
  {\rm Im}I_2(k ^ 0 + i \epsilon  ,\vec{k})
  &=
  \frac{{\rm sgn}(k^0)}{2(4\pi)^3}
  \mu  ^{ - 4 \varepsilon  } 
  \left(\frac{1}{\varepsilon}
    +2\ln \frac{\bar{\mu}^2}{k^2}+4\right)+\ord (\varepsilon)
  ,\\
  {\rm Im}I_3(k+i \epsilon,\vec{k})
  &=
  -  
  \frac{{\rm sgn}(k^0)k^2}{8(4\pi)^3} 
  +\ord (\varepsilon)
  .
\end{align}
This yields the results (\ref{rho3L})-(\ref{rho4R}).

\subsection{Non-factorizable integrals}

In the more general case that the three-point correlator cannot be
written as a product of two two-point correlators, we proceed as
follows.  We consider  the $N_L$-loop Feynman integrals   with $N_p$
propagators
\begin{equation}
  I(k_1,k_2)
  =
  \left(\prod_{l=1}^{N_L} 
    \int \frac{d^d p_l}{(2\pi)^d} \right)
  \prod_{i=1}^{N_p} \frac{1}{q_i^2}
  ,\qquad 
  q_i=a_{in} p_n+b_{im}k_{m},
\end{equation}
with real coefficients $a_{in}$ and $b_{im}$.  In order to compute the
spectral function $\rho(-k,k)$ we apply the following steps.
\begin{enumerate}
\item Compute the integrals over $ p_i ^ 0$. The result can be written
  in the form
 \begin{align}
   I(k_1,k_2)
   =&
   \left(\prod_{l=1}^{N_L} \int \frac{d^{d-1} p_l}{(2\pi)^{d-1}} \right)
   \frac{A(\vec{k}_i,\vec{p}_i)}
   {(B(\vec{k}_i,\vec{p}_i)- k _ 1 ^ 0)
     (C(\vec{k}_i,\vec{p}_i)- k _ 2 ^ 0)}
   \nonumber\\
   +&\text{ many similar terms}
\end{align}
Here $A$, $B$ and $C$ are real functions which only depend on scalar products of 
the spatial components of the external momenta
and the loop momenta.
\item Apply the inverse relation (\ref{invsimple}) and
  (\ref{deltadisc}). The result yields two delta-functions such that  
  \begin{align}
    \rho(k_1,k_2)
    =&
    4\pi^2\left(\prod_{l=1}^{N_L} 
      \int \frac{d^{d-1} p_l}{(2\pi)^{d-1}} \right)
    A(\vec{k}_i,\vec{p}_i) 
    \delta(B(\vec{k}_i,\vec{p}_i)-k_1^0)
    \delta(C(\vec{k}_i,\vec{p}_i)-k_2^0)
    \nonumber\\
    +&\text{ many similar terms}
    .
\end{align}
\item Set $k_2=-k_1=k$ with $k^2=M_1^2$ and drop all terms which do
  not contribute due to the constraints of the delta functions.
\item Solve the remaining $(d-1)$-dimensional integrals.
\item Expand the result in $\varepsilon=(d-4)/2$.
\end{enumerate}
This yields the results (\ref{rhoBB}),(\ref{rhoLR})-(\ref{rho4R}).

If one propagator is squared, that is,
\begin{equation}
   I_{\rm squared}(k_1,k_2)
   =
   \left(\prod_{l=1}^{N_L} \int \frac{d^d p_l}{(2\pi)^d} \right)
   \frac{1}{q_k^4}\prod_{i\neq k}^{N_p} \frac{1}{q_i^2}
   ,
   \label{squard}
\end{equation}
we introduce an artificial mass as
\begin{equation}
  I_m(k_1,k_2)
  =
  \left(\prod_{l=1}^{N_L} 
    \int \frac{d^d p_l}{(2\pi)^d} \right)\frac{1}{q_k^2-m^2}
  \prod_{i\neq k}^{N_p} \frac{1}{q_i^2}
  .
\end{equation}
Then we apply the steps (1.) - (5.) for $I_m$ and obtain the spectral
function $\rho_{\rm squared}$ as
\begin{equation}
  \rho_{\rm squared}(-k,k)
  =
  \left . 
  \frac{d^2}{dm^2} \rho_m(-k,k)
  \right | _ { m ^ 2 = 0 } 
  ,
\end{equation}
which yields the result (\ref{rhoBBdot}).

\section{Treatment of $\gamma^5$ in Dirac traces}\label{Dirac}

In this appendix we argue that terms with $\gamma^5$ do not contribute
to the Dirac traces.  Contracting gauge indices and using the
properties of the charge conjugation
matrix $\cmatrix$ the NLO diagrams read in Feynman gauge
\begin{align}
  \TopGlassesD(\LNb,\Ahh,\Aqb,\Ahh,\Ahh,\Ahh,\Aq,\Agl,\Lge)
  &=
  2\Nweak(\Nweak+1)(y^2_{\tilde{\varphi}}g_1^2+C_2(r)g_2^2)
  {\rm Tr}\left( \gamma_{\mu_1}P_{\rm L} \gamma_{\mu_2} 
    P_{\rm R}\right)
  \nonumber\\ 
  &\times
  \int_{p_1,p_2,p_3} \frac{p_1^{\mu_1}p_3^{\mu_2} ( p_1 + p_2-2k_1)^2  }
     { p_1^2 p_3^2 (p_1-k_1)^4 (p_2-k_1)^2(p_3+k_2)^2 (p_1-p_2)^2}
     ,
     \label{gfac1}\\[0.4cm]
     \TopGlassesC(\LNb,\Ahh,\Aq,\Ahh,\Ahh,\Ahh,\Aqb,\Agl,\Lge)
     &=2
     \Nweak(\Nweak+1)(y^2_{\tilde{\varphi}}g_1^2+C_2(r)g_2^2)
     {\rm Tr}\left( \gamma_{\mu_1}P_{\rm L} \gamma_{\mu_2} 
       P_{\rm R}\right)
     \nonumber\\ 
     &
     \times\int_{p_1,p_2,p_3}
     \frac{p_1^{\mu_1}p_2^{\mu_2} ( p_3 +p_2 + 2k_2 )^2  }
     {p_1^2 p_2^2 (p_1-k_1)^2  (p_2+k_2)^4(p_3+k_2)^2 (p_3-p_2)^2}
     ,
     \label{gfac2}\\[0.4cm]
     \TopGlassesE(\LNb,\Ahh,\Aqb,\Aq,\Aq,\Aq,\Ahh,\Agl,\Lge)
     &=2
     \Nweak(\Nweak+1)(y^2_{\ell}g_1^2+C_2(r)g_2^2)
     \nonumber\\
     &
     \times{\rm Tr}\left(\gamma_{\mu_1} P_{\rm L} \gamma_{\mu_5} 
       P_{\rm R}\gamma_{\mu_2} P_{\rm L} \gamma^{\mu_5} 
       P_{\rm R}\gamma_{\mu_3} P_{\rm L}\gamma_{\mu_4} P_{\rm R}\right)
     \nonumber\\
     &
     \times\int_{p_1,p_2,p_3}
     \frac{p_1^{\mu_1}p_2^{\mu_2}p_1^{\mu_3}p_3^{\mu_4}  }
     {p_1^4 p_2^2p_3^2 (p_1-k_1)^2 (p_3+k_2)^2 (p_1-p_2)^2  }
     ,\label{gfac3}\\[0.4cm]
     \TopGlassesF(\LNb,\Ahh,\Aq,\Aqb,\Aqb,\Aqb,\Ahh,\Agl,\Lge)
     &=
     2\Nweak(\Nweak+1)(y^2_{\ell}g_1^2+C_2(r)g_2^2)
     \nonumber\\
     &
     \times{\rm Tr}\left( \gamma_{\mu_1}P_{\rm L}\gamma_{\mu_2}
       P_{\rm R}\gamma^{\mu_5}P_{\rm L}\gamma_{\mu_3}P_{\rm R}\gamma_{\mu_5}
       P_{\rm L}\gamma_{\mu_4}P_{\rm R}\right)
     \nonumber\\
     &
     \times\int_{p_1,p_2,p_3}
     \frac{  p_1^{\mu_1}p_2^{\mu_2}p_3^{\mu_3}p_2^{\mu_4}  }
     {p_1^2 p_2^4 p_3^2 (p_1-k_1)^2 (p_2+k_2)^2 (p_2-p_3)^2}
     ,
     \label{gfac4}\\[0.4cm]
     \TopGlassesA(\LNb,\Ahh,\Ahh,\Aq,\Aq,\Ahh,\Aqb,\Lgl,\Lge)
     &=
     2\Nweak(\Nweak+1)(y_{\tilde{\varphi}}y_{\ell}g_1^2+C_2(r)g_2^2)
     \\
     &\times
     {\rm Tr}\left( \gamma_{\mu_1}P_{\rm L}\gamma_{\mu_5}P_{\rm R}
       \gamma_{\mu_2}P_{\rm L}\gamma_{\mu_3} P_{\rm R} 
     \right)
     \nonumber \\
     &
     \times\int_{p_1,p_2,p_3}
     \frac{p_1^{\mu_1}p_2^{\mu_2}p_3^{\mu_3}(p_1 +p_2-2k_1)^{\mu_5}  }
     {p_1^2 p_2^2 p_3^2(p_1-k_1)^2 (p_2-k_1)^2 (p_3+k_2)^2 (p_1-p_2)^2}
     ,
     \label{gfac5}\\[0.4cm]
     \TopGlassesB(\LNb,\Ahh,\Aq,\Ahh,\Ahh,\Aqb,\Aqb,\Lgl,\Lge)
     &=
     2\Nweak(\Nweak+1)(y_{\tilde{\varphi}}y_{\ell}g_1^2+C_2(r)g_2^2)
     \\
     &\times
     {\rm Tr}\left(\gamma_{\mu_1}P_{\rm L}\gamma_{\mu_2}P_{\rm R}
       \gamma_{\mu_5}P_{\rm L}\gamma_{\mu_3} P_{\rm R}
     \right)
     \nonumber\\
     &
     \times\int_{p_1,p_2,p_3}
     \frac{ p_1^{\mu_1}p_2^{\mu_2}p_3^{\mu_3} (p_2 +p_3 + 2k_2 )^{\mu_5}  }
     { p_1^2 p_2^2 p_3^2 (p_1-k_1)^2 (p_2+k_2)^2 (p_3+k_2)^2 (p_2-p_3)^2}
     ,
     \label{gfac6}
\end{align}
\begin{align}
  \TopGlassesC(\LNb,\Ahh,\Aq,\Ahh,\Aq,\Ahh,\Aqb,\Aq,\Lge)
  &=
  -2\Nweak(\Nweak+1)N_c|h_t|^2
  \\ &
  \times{\rm Tr}\left( \gamma_{\mu_1}P_{\rm L}\gamma_{\mu_2}P_{\rm R}\right) 
  {\rm Tr}\left(\gamma_{\mu_3} P_{\rm L} \gamma_{\mu_4} P_{\rm R} 
  \right)
  \nonumber\\ &
  \times\int_{p_1,p_2,p_3}
  \frac{p_1^{\mu_1}p_2^{\mu_2} (p_3 + k_2)^{\mu_3}( p_2 - p_3 )^{\mu_4}}
  {p_1^2 p_2^2 (p_1-k_1)^2  (p_2+k_2)^4(p_3+k_2)^2 (p_3-p_2)^2}
  ,
  \label{top1}\\[0.4cm]
  \TopGlassesD(\LNb,\Ahh,\Aqb,\Ahh,\Aq,\Ahh,\Aq,\Aq,\Lge)
  &=
  -2\Nweak(\Nweak+1)N_c|h_t|^2
  \\
  &
  \times{\rm Tr}\left(\gamma_{\mu_1} P_{\rm L}\gamma_{\mu_2} P_{\rm R}\right) 
  {\rm Tr}\left(\gamma_{\mu_3} P_{\rm L}\gamma_{\mu_4}P_{\rm R} 
  \right)
  \nonumber\\  &
  \times\int_{p_1,p_2,p_3}
  \frac{ p_1^{\mu_1}p_3^{\mu_2} (k_1-p_2)^{\mu_3}(p_2 - p_1 )^{\mu_4} }
  { p_1^2 p_3^2 (p_1-k_1)^4 (p_2-k_1)^2(p_3+k_2)^2 (p_1-p_2)^2}
  ,
  \label{top2}
\end{align}
\begin{align}
  \TopScaryA(\LNb,\Aq,\Aqb,\Ahh,\Ahh,\Ahh,\Ahh,\Lge)
  =&
  4\Nweak(\Nweak+1)\lambda{\rm Tr}
  \left(\gamma_{\mu_1} P_{\rm L}\gamma_{\mu_2}P_{\rm R}\right)
  \nonumber\\&
  \times\int_{p_1, p_2, p_3}\frac{(k_1-p_1)^{\mu_1}(p_2+k_2)^{\mu_2}}
  {p_1^2p_2^2(p_1-k_1)^2(p_2+k_2)^2(p_1-p_3)^2(p_2-p_3)^2}
  ,
  \label{Hi}
\end{align}
\begin{align}
  \TopScaryB(\LNb,\Aq,\Aqb,\Ahh,\Ahh,\Agl,\Lhh,\Lhh,\Lge)&
  =
  2\Nweak(y_{\tilde{\varphi}}^2(\Nweak+1)g_1^2+C_2(r)g_2^2)
  {\rm Tr}\left(\gamma_{\mu_1}P_{\rm L}\gamma_{\mu_2}P_{\rm R}\right)
  \nonumber\\
  &\times\int_{p_1,p_2,p_3}
  \frac{(k_1-p_1)^{\mu_1} (p_2+k_2)^{\mu_2}(p_3-2p_2)^{\mu_5}(p_3-2p_1)_{\mu_5}}
  {p_1^2p_2^2p_3^2(p_1-k_1)^2(p_2+k_2)^2(p_1-p_3)^2(p_2-p_3)^2}
  ,\label{gV1}\\[0.4cm]
  \TopScaryB(\LNb,\Ahh,\Ahh,\Aq,\Aqb,\Agl,\Lqb,\Lqb,\Lge)
  &=
  -2\Nweak(y^2_{\ell}(\Nweak+1)g_1^2+C_2(r)g_2^2)
  \nonumber\\
  &
  \times{\rm Tr}
  \left(\gamma_{\mu_1}P_{\rm L}\gamma_{\mu_5}P_{\rm R}
    \gamma_{\mu_2}P_{\rm L}\gamma_{\mu_3}P_{\rm R}\gamma^{\mu_5}
    P_{\rm L}\gamma_{\mu_4}P_{\rm R}\right)
  \nonumber\\ &
  \times\int_{p_1,p_2,p_3}
  \frac{p_1^{\mu_1}(p_1 - p_3)^{\mu_2}(p_2-p_3)^{\mu_3}p_2^{\mu_4}}
  {p_1^2p_2^2p_3^2(p_1-k_1)^2(p_2 + k_2)^2(p_1-p_3)^2(p_2-p_3)^2}
  ,
  \label{gV2}\\[0.4cm]
  \TopScaryB(\LNb,\Ahh,\Aqb,\Ahh,\Aqb,\Agl,\Lhh,\Lqb,\Lge)
  &=
  2\Nweak(y_{\tilde{\varphi}}y_{\ell}(\Nweak+1)g_1^2+C_2(r)g_2^2)
  \\
  &
  \times{\rm Tr}\left(\gamma_{\mu_1}P_{\rm L}\gamma_{\mu_2}
    P_{\rm R}\gamma_{\mu_5}P_{\rm L}\gamma_{\mu_3}P_{\rm R}\right)
  \nonumber\\
  &
  \times\int_{p_1,p_2,p_3}
  \frac{ (k_1-p_1)^{\mu_1}( p_3 - p_1)^{\mu_2}p_2^{\mu_3} (p_3-2p_1)^{\mu_5}}
  {p_1^2p_2^2p_3^2(p_1-k_1)^2(p_2+k_2)^2(p_1-p_3)^2(p_2-p_3)^2}
  ,
  \label{gV3}\\[0.4cm]
  \TopScaryB(\LNb,\Aq,\Ahh,\Aq,\Ahh,\Agl,\Lqb,\Lhh,\Lge)
  &=
  2\Nweak(y_{\tilde{\varphi}}y_{\ell}(\Nweak+1)g_1^2+C_2(r)g_2^2)
  \\
  &
  \times{\rm Tr}\left(\gamma_{\mu_1}P_{\rm L}\gamma_{\mu_5}
    P_{\rm R}\gamma_{\mu_2}P_{\rm L}\gamma_{\mu_3}P_{\rm R} 
  \right)
  \nonumber\\ &
  \times\int_{p_1,p_2,p_3}
  \frac{ p_1^{\mu_1}(p_1 -p_3)^{\mu_2}(p_2 + q)^{\mu_3}(p_3-2p_2)^{\mu_5}}
  {p_1^2p_2^2p_3^2(p_1-k_1)^2(p_2+k_2)^2(p_1-p_3)^2(p_2-p_3)^2}
  .
  \label{gV4}
\end{align}
Here we use $y_{\varphi}=-y_{\tilde{\varphi}}=-y_{\ell}=1/2$,
$\Nweak=2$, $C_2(r)=3/4$ and $N_c=3$.

The structure of the traces in the factorizable diagrams
(\ref{gfac1})-(\ref{top2}) has been already analyzed in
\cite{Laine:2011pq}, where it has been found that the $\gamma^5$
contributions drop out.  The non-factorizable diagrams (\ref{Hi}) and
(\ref{gV1}) contain the same trace as in the leading order case, so
that $\gamma^5$ drops out there as well, but for the diagrams
(\ref{gV2})-(\ref{gV4}) we need a further argument.  Using a naively
anti-commuting $\gamma^5$ and $(\gamma^5)^2=1$ in traces with more than
one $\gamma^5$, only traces with no or one $\gamma^5$ remain.  With
the definition (\ref{gam5}) we can write these traces as
\begin{align}
 {\rm Tr}\left(\gamma_{\mu_1}\gamma_{\mu_2}\gamma_{\mu_3}\gamma_{\mu_4}
   \gamma^5\right)
 &= -
 \frac{i}{4!}\epsilon^{\nu_1\nu_2\nu_3\nu_4}
 {\rm Tr}\left(\gamma_{\mu_1}\gamma_{\mu_2}
   \gamma_{\mu_3}\gamma_{\mu_4}\gamma_{\nu_1}
   \gamma_{\nu_2}\gamma_{\nu_3}\gamma_{\nu_4}\right).
\end{align}
The total anti-symmetry of the Levi-Civita symbol $
\epsilon^{\nu_1\nu_2\nu_3\nu_4} $ guarantees that only totally
anti-symmetric combinations of $\eta^{\mu_1\nu_1} \eta^{\mu_2\nu_2}
\eta^{\mu_3\nu_3} \eta^{\mu_4\nu_4}$ contribute to the trace, which
yields
\begin{equation}
  {\rm Tr}\left(\slashed k_1 \slashed k_2\slashed k_3\slashed k_4\gamma^5\right)
  = -
  4i\epsilon_{\mu\nu\rho\sigma} k_1^{[\mu} k_2^{\nu} k_3^{\rho} k_4^{\sigma]}
  .
\end{equation}
Therefore, all tensor integrals in the diagrams (\ref{gV2}) -
(\ref{gV4}) coming from terms with traces containing one $\gamma^5$,
have the generic form
\begin{equation}
 I_T^{\mu\nu\rho\sigma}(k_1,k_2)=\int_{p_1,p_2,p_3}
 \frac{T^{\mu\nu\rho\sigma}(\{p_{i}\},k_1,k_2)}
 {p_1^2p_2^2p_3^2(p_1-k_1)^2(p_2+k_2)^2(p_1-p_3)^2(p_2-p_3)^2}
 ,
 \label{IT}
\end{equation}
where $T^{\mu_1\mu_2\mu_3\mu_4}(\{p_{i}\},k_1,k_2)$ is one of the six
total anti-symmetric rank-4-tensors of the set
\begin{equation}
  \{p_1^{[\mu} p_2^{\nu} p_3^{\rho} k_i^{\sigma]},p_i^{[\mu} p_j^{\nu} 
  k_1^{\rho} k_2^{\sigma]}\}
  . 
\end{equation}
Some of the tensor integrals of the class (\ref{IT}) contain the vector-integral
\begin{equation}
  J^{\mu}(p_1,p_2)=\int_{p_3}\frac{p_3^{\mu}}{p_3^2(p_3-p_1)^2(p_3-p_2)^2}
\end{equation}
as a sub-integral.  Due to Lorentz invariance it can be written in
terms of scalar functions $f_1$ and $f_2$ as
\begin{equation}
  J^{\mu}(p_1,p_2)=p_1^\mu f_1(p_1,p_2)+ p_2^\mu f_2(p_1,p_2)
  .
\end{equation}
Therefore, we can express all integrals of the class (\ref{IT}) in terms of 
integrals containing only the tensor
\begin{equation}
  T^{\mu\nu\rho\sigma}(p_1,p_2,k_1,k_2)=p_1^{[\mu} p_2^{\nu} k_1^{\rho}
  k_2^{\sigma]}
  . 
\end{equation}
Lorentz symmetry allows to compute the three-point correlator for
$k_1=(k_1^0,\vec{0})$ and $k_2=(k_2^0,\vec{0})$.  Since
$T^{\mu\nu00}(p_1,p_2,k_1,k_2)=0$, terms with $\gamma^5$ do not
contribute.  This argument holds only at zero temperature. For a
finite-temperature computation it would be necessary to compute the tensor
sum-integrals explicitly.





\end{document}